\documentstyle[12pt,epsfig]{article}
\newcommand{\slsh}[1]{\not{\hbox{\kern-2pt${#1}$}}}

\newcommand{\ba}[1]{\begin{eqnarray} \label{#1}}
\newcommand{\ea}{\end{eqnarray}}

\def\beq{\begin{equation}}
\def\eeq{\end{equation}}
\def\bea{\begin{eqnarray}}
\def\eea{\end{eqnarray}}
\def\bqu{\begin{quote}}
\def\equ{\end{quote}}

\def\gappeq{\mathrel{\rlap {\raise.5ex\hbox{$>$}}
{\lower.5ex\hbox{$\sim$}}}}

\def\lappeq{\mathrel{\rlap{\raise.5ex\hbox{$<$}}
{\lower.5ex\hbox{$\sim$}}}}

\begin{document}
\pagestyle{empty}
\begin{flushright}
{CERN-TH/2008-225}\\
{UHU-GEM/21-2008}\\
\end{flushright}

\vspace*{1 cm}
\begin{center}
{\large {\bf Search for Tau Flavour Violation at the LHC}} \\
\vspace*{1cm}
{\bf E. Carquin$^1$, J.~Ellis$^2$, M.E. G{\'o}mez$^3$, S.~Lola$^4$
and
{\bf J.Rodriguez-Quintero$^3$}\\
}
\vspace{0.3cm}
$^1$ Centre of Subatomic Studies, Technical University Federico Santa Mar\'ia,
Valpara\'iso, Chile \\
$^2$ Theory Division, Physics Department, CERN, CH-1211 Geneva 23, Switzerland \\
$^3$ Department of Applied Physics, University of Huelva, 21071 Huelva, Spain \\
$^4$ Department of Physics, University of Patras, 26500 Patras, Greece 

\vspace*{2cm}
{\bf ABSTRACT} \\ 
\end{center}
\vspace*{5mm}
\noindent

We explore the prospects for searches  at the LHC for 
sparticle decays that violate $\tau$ lepton number,
in the light of neutrino oscillation data and the seesaw model for neutrino masses and mixing.
We analyse the theoretical and phenomenological conditions required for tau flavour violation
to be observable in $\chi_2 \to \chi +  \tau^\pm \mu^\mp$ decays, 
for cosmologically interesting values of
the relic neutralino LSP density. We study the 
relevant supersymmetric parameter space in
the context of the Constrained Minimal Supersymmetric Extension 
of the Standard Model (CMSSM) and in SU(5) extensions of the theory.
We pay particular attention to the possible
signals from hadronic tau decays, that we  
analyse using {\tt PYTHIA} event simulation. We find that a signal for $\tau$ flavour-violating
$\chi_2$ decays may be observable if the branching ratio exceeds about 10\%.
This may be compatible with the existing upper limit on $\tau \to \mu \gamma$ decays if
there is mixing between right-handed sleptons, as could be induced in non-minimal SU(5) GUTs.

\vspace*{5cm}
\noindent

\setcounter{page}{1}
\pagestyle{plain}

\section{Introduction}

Data from  both atmospheric~\cite{skatm} and solar~\cite{sksol}
neutrinos have confirmed the existence of neutrino oscillations
with near-maximal  $\nu_\mu - \nu_\tau$ mixing (Super-Kamiokande) and
large $\nu_e \to \nu_{\mu}$ mixing (SNO).
These observations would also imply violation of the
corresponding charged-lepton numbers, which would be enhanced 
in supersymmetric theories and might be observable in low-energy
experiments. In fact, charged-lepton-number violating processes could occur
at embarrassingly large rates if the soft supersymmetry-breaking masses of
the squarks and sleptons were not universal. For this reason, it is often
assumed that these masses are equal at the grand-unification scale, as in
the constrained minimal supersymmetric extension of the Standard Model
(CMSSM).

Even within  the minimal supersymmetric version of the
seesaw model for neutrino masses, renormalization of the soft 
supersymmetry-breaking slepton masses would occur 
while running from the GUT scale to the heavy neutrino mass scales.
This would be induced by the Dirac Yukawa couplings of
the neutrinos~\cite{bm}, since these cannot, in general, be diagonalised simultaneously
with the charged-lepton and the slepton mass matrices.
This scenario provides a minimal amount of charged-lepton-flavour 
violation, which could be further enhanced
by GUT interactions and/or non-universal slepton masses at the GUT scale.

Within this framework, many signatures for charged-lepton-flavour 
violation have been considered ~\cite{rev,LFVres}, 
including $\mu \to e \gamma$  decays, $\mu-e$ conversions,
$\tau \to \mu \gamma$ and $\tau \to e
\gamma$ decays. In view of the (near-)maximal
mixing observed amongst the corresponding neutrino species,
one expects these decays to be relatively large when the soft
supersymmetry-breaking masses $M_{1/2}$ and/or $m_0$ are relatively small. Other charged-lepton-flavour 
violating possibilities that have been considered are the decays 
$\chi_2\to \chi + e^\pm \mu^\mp$ \cite{sleptonoscemu,Hisano}, 
and $\chi_2\to \chi + \mu^\pm \tau^\mp$ \cite{HP,CEGLR}, 
where $\chi$ is the lightest
neutralino, assumed here to be the lightest supersymmetric particle (LSP),
and $\chi_2$ is the second-lightest neutralino. It has been argued that
these decays might have a rate observable at the LHC for certain choices of
the CMSSM parameters~\cite{HP,CEGLR}. These decays would provide search possibilities
that are complementary to searches for the flavour-violating decays
of charged leptons, since they may be relevant
particularly for regions of the supersymmetric parameter space where rare
charged lepton decays and conversions are suppressed.

The answer to the question which decay mode offers better detection prospects
at the LHC or at a future linear collider depends
on the details of both the theoretical model and the experiment.
The decay  $\chi_2 \to \chi +\mu^\pm \tau^\mp$ 
has an experimental signature that is less distinctive than $\chi_2 \to \chi + e^\pm \mu^\mp$. 
However, it may have certain theoretical
advantages over the latter mode. This is because
$\nu_\tau - \nu_\mu$ mixing is known to be essentially maximal, and
the feedthrough into the charged-lepton sector is potentially
enhanced by larger Dirac Yukawa couplings and/or lighter singlet-neutrino
masses, if neutrino masses exhibit the expected hierarchical pattern.
These comments imply that the two lepton-flavour-violating modes are complementary,
and both have to be studied. In order to assess the observability of
$\chi_2 \to \chi +\tau^\pm \mu^\mp$, the detailed simulation of
signal and background events seems unavoidable.

In previous work~\cite{CEGLR}, we observed that the branching ratio 
for $\chi_2 \to \chi + \tau^\pm \mu^\mp$
decay is enhanced when $m_{\chi_2} > m_{\tilde \tau_1} > m_\chi$, where
${\tilde \tau_1}$ is the lighter stau slepton. This occurs in a wedge of
the $(M_{1/2}, m_0)$ parameter plane in the CMSSM that is complementary to
that explored by $\tau \to \mu \gamma$.
The region of CMSSM parameter
space where this enhancement occurs includes the region where $\chi -
{\tilde \ell}$ coannihilation suppresses the relic density $\Omega_\chi$,
keeping it within the range $\Omega_\chi h^2 \sim 0.1$ preferred by
astrophysics and cosmology, even if $M_{1/2}$ is comparatively large.
The interest of this coannihilation region is supported by 
experimental constraints on the CMSSM, such as $m_h$ and $b \to s \gamma$
decay, which disfavour low values of $M_{1/2}$.  

In the current paper, we revisit the $\chi_2 \to \chi + \tau^\pm \mu^\mp$
decay mode and extend previous analyses
in the following directions:

$\bullet$
{\em Theoretical framework}: In \cite{CEGLR} we considered 
generic seesaw mixing within the framework of the CMMSM.
Here, we will study how the ratio
$\Gamma(\chi_2\rightarrow\chi+ \tau^\pm+ \mu^\mp)/
\Gamma(\chi_2\rightarrow\chi+ \tau^\pm+ \tau^\mp)$
is modified under different theoretical assumptions,
and what information a possible
signal may provide in this respect.
Specifically, in our previous analysis we
considered only the dominant effects due to large mixing in the 
2-3 sector of the  charged-slepton mass 
matrix.  Within this framework,  the slepton mixing
arises essentially only from the LL sector, and
large LFV effects may be observed mostly  
at low $\tan\beta$, where the smuon and stau are almost degenerate.
As $\tan\beta$ increases, the LFV width decreases 
if flavor is only violated in the LL sector,
since the lightest stau becomes mostly right-handed.
However, in a GUT-inspired model
the running of couplings from the Planck to the GUT scale may 
introduce significant corrections to the right-handed slepton masses,
giving in principle rise to the possibility of enhanced rates at
large $\tan\beta$ as well. We address these considerations
in a more elaborate study of 
the complete mixing effects.

$\bullet$
{\em Detailed study of the supersymmetric spectrum and parameter space}: 
We pay particular attention to regions that lead
to large values of $\Gamma(\chi_2\rightarrow\chi+ \tau^\pm+ \tau^\mp)$ 
through on-shell slepton production,
($BR(\chi_2\rightarrow\chi \tau^\pm \mu^\mp)=\sum_{i=1}^3 \left[
BR(\chi_2\rightarrow\tilde{l}_i \mu)BR(\tilde{l}_i\rightarrow\tau \chi) + 
BR(\chi_2\rightarrow\tilde{l}_i \tau)BR(\tilde{l}_i\rightarrow\mu \chi)
\right]$),
while satisfying all phenomenological
and  cosmological (relic density) constraints. 
The characteristic parameter region for
the signal in the $\tau$ channel to be
optimal is defined by the following:\\
(i) $m_{\chi_2}> m_{\tilde{\tau}}>m_\chi$;\\
(ii) we also assume that one of the mass differences in (i) is $> m_\tau$
and the other $>m_\mu$, $m_{\tilde{\tau}} > m_\chi$, so that the $\mu, \tau$
and $\tilde{\tau}$ are all on-shell;\\
(iii) Moderate values of $m_\chi$ (phase space and luminosity considerations). \\
These conditions are obeyed in significant fractions
of the stau coannihilation region.


$\bullet$
{\em {\tt PYTHIA} event simulation}:  We use {\tt {\tt PYTHIA}} to simulate the
hadronic decays of $\tau$s produced in the dilepton decay of the second-lightest 
neutralino, $\chi_2^0 \rightarrow\tilde{l}l\rightarrow\chi ll$~\cite{HP}. 
In the study of the flavor-violating dilepton signal ($\tau^{\pm}\mu^{\mp}$),
the second lepton is tagged as a muon 
with a probability equal to the branching ratio assumed for flavor-violating decays.


After first setting the theoretical scene, our procedure in this paper is a bottom-up one, namely: \\
(i) Knowing what branching ratio would be required 
for an observable signal at the LHC, through an event simulation performed using {\tt PYTHIA};\\
(ii) We study the theoretical frameworks that may satisfy the conditions for
observability, together with the supersymmetric parameter space requirements. 
In doing so, we review look at models with mixing arising mainly in the 
left-handed slepton sector, which could not produce an observable signal,
and then discuss the possibility of mixing in the 
right-handed sector, which becomes possible in non-minimal GUTs, where 
effects due to renormalization above $M_{GUT}$ schemes may
become relevant. 

The structure of our paper is as follows. Different theoretical frameworks are reviewed in 
Section 2, the supersymmetric
parameter space is explored in Section~3, our simulation is described in Section~4
and its results in Section~5. Our conclusions are summarized in Section~6.

\section{Possible Sources of Charged-Lepton Flavour Violation}

\subsection{Renormalization below $M_{GUT}$}
\label{minSU5}

In the minimal supersymmetric extension of the seesaw mechanism for 
generating neutrino masses with three heavy 
singlet-neutrino states $N_i$, the leptonic sector of the 
superpotential is:
\begin{equation}
W = N^{c}_i (Y_\nu)_{ij} L_j H_2
  -  E^{c}_i (Y_e)_{ij}  L_j H_1
  + \frac{1}{2}{N^c}_i {\cal M}_{ij} N^c_j + \mu H_2 H_1 \,,
\label{leptonW}
\end{equation}
where $Y_\nu$ is the neutrino Dirac Yukawa coupling matrix, ${\cal 
M}_{ij}$ is the Majorana mass matrix for the $N_i$, the $L_j$ and $H_I$ 
are lepton and Higgs doublets, and the $E^{c}_i$ are singlet 
charged-lepton supermultiplets. 
The superpotential of  the effective 
low-energy theory, obtained after the decoupling of 
heavy neutrinos is  \cite{valle}
\begin{eqnarray}
\label{weff}
W_{eff}&=& L_i H_2 \left(Y_\nu^T \left({\cal M}^{D}\right)^{-1} 
Y_\nu\right)_{ij} L_j H_2 
  -  E^{c}_i (Y_e)_{ij}  L_j H_1.
\label{LEET}
\end{eqnarray}
In the basis where the charged leptons and
the heavy neutrino mass matrices are diagonal, one finds
\bea 
{\cal M}_\nu={Y}_\nu^T \left({\cal M}^{D}\right)^{-1} {
Y}_\nu v^2 \sin^2\beta ,
\label{seesaw1}
\eea
where  $v=174$ GeV and  $\tan\beta \equiv v_2/v_1$.

In the context of the CMSSM, the soft 
supersymmetry-breaking masses of the charged and neutral sleptons are 
assumed to be universal at the GUT scale, with a common value $m_0$. In 
the leading-logarithmic approximation, the non-universal renormalization 
of the soft supersymmetry-breaking scalar masses is given by 
\begin{eqnarray}
(m^2_{\tilde{L}})_{ij} 
&\simeq& - \frac{1}{8\pi^2} \left( 
  \lambda_{\nu_3}^2 V^{\ast 3i}_{D} V_{D}^{3j} 
 \log \frac{M_{\rm grav}}{M_{\nu_3}}
+ \lambda_{\nu_2}^2 V^{\ast 2i}_{D} V_{D}^{2j} 
 \log \frac{M_{\rm grav}}{M_{\nu_2}}
\right)  (3 m_0^2 +a^2_0)  .
\label{offdiagofL} 
\end{eqnarray}
implying that the corresponding corrections to left-handed slepton masses
are given by $V_{D}$, the Dirac neutrino mixing matrix in the basis where
the $d$-quark and charged-lepton masses are diagonal.
In this approach, non-universality in the soft 
supersymmetry-breaking left-slepton masses is much larger than that in the
right-slepton masses, particularly when the trilinear soft 
supersymmetry-breaking parameter $A_0 = 0$. 
Consequently, within the CMSSM,
to a good approximation,
the renormalization of the soft supersymmetry-breaking parameters at low 
energies can be understood  in terms of the dominant 
non-universality in the third-generation left-slepton mass:
$m^2_{0_{LL}} =  {\rm diag} (m_0^2, m_0^2, x \times m_0^2)$,
where a typical value of the non-universality factor is $x \sim 0.9$. 
Based on the above, in \cite{CEGLR} we assumed that
there is an off-diagonal ${\tilde \tau_L} - 
{\tilde 
\mu_L}$ mixing term in the soft mass-squared matrix
\begin{equation}
\Delta m^2_{0_{LL}} \sim (1-x) m^2_0 {\sin (2\phi) \over 2},
\label{offdiag}
\end{equation}
where $\phi$ is the mixing angle between the second and third generation
in the charged-lepton Yukawa matrix. This mixing leads to lepton-flavour
violation $\sim \sin^2 (2\phi)$.

In our current work, we go beyond the above approximation by including
the complete  mixing effects in interesting regions of the parameter space.
For the structures of the mixing matrices, as well as the heavy and light neutrino
hierarchies, one must appeal to a 
specific GUT model. Here we consider textures obtained by 
combining SU(5) with a U(1) family symmetry.
Requiring that fields in the same GUT multiplets have the
same flavour charge leads straightforwardly to
fermion mass matrices of the following forms~\cite{SU5-a,SU5-b}
 \begin{equation}
{\cal M}_{u}\propto  \left( 
\begin{array}{ccc}
\bar{\epsilon}^{6} & \bar{\epsilon}^{5} & \bar{\epsilon}^{3} \\ 
\bar{\epsilon}^{5} & \bar{\epsilon}^{4} & \bar{\epsilon}^{2} \\ 
\bar{\epsilon}^{3} & \bar{\epsilon}^{2} & 1
\end{array}
\right) ,
~~M_{down}\propto 
\left( 
\begin{array}{ccc}
\bar{\epsilon}^{4} & \bar{\epsilon}^{3} & \bar{\epsilon}^{3} \\ 
\bar{\epsilon}^{3} & \bar{\epsilon}^{2} & \bar{\epsilon}^{2} \\ 
\bar{\epsilon} & 1 & 1
\end{array}
\right) ,
~~M_{\ell }\propto  \left( 
\begin{array}{ccc}
\bar{\epsilon}^{4} & \bar{\epsilon}^{3} & \bar{\epsilon} \\ 
\bar{\epsilon}^{3} & \bar{\epsilon}^{2} & 1 \\ 
\bar{\epsilon}^{3} & \bar{\epsilon}^{2} & 1
\end{array}
\right) ,
\label{EX1}
\end{equation}
with $\bar{\epsilon}\sim 0.2$ as the preferred value.
The mass structure and the mixings
of the neutrinos are more complicated, because of the heavy Majorana masses of the
right-handed components arising
from terms of the form
$\nu_R\nu_R\Sigma$, where $\Sigma$ is an SU(3)$\times$
SU(2)$\times$U(1)-invariant Higgs field with $I_W=0$ and a non-zero flavour U(1) 
charge. A working example was given in~\cite{EGL}: 
\begin{eqnarray}
M_{RR}& \propto & \left[
\begin{array}{ccc}
\bar{\epsilon}^{|2n_1+\sigma|}&\bar{\epsilon}^{|n_1+n_2+\sigma|}&\bar{\epsilon}^{|n_1+n_3+\sigma|}\\
\bar{\epsilon}^{|n_1+n_2+\sigma|}&\bar{\epsilon}^{|2n_2+\sigma|}&\bar{\epsilon}^{|n_2+n_3+\sigma|}\\
\bar{\epsilon}^{|n_1+n_3+\sigma|}&\bar{\epsilon}^{|n_2+n_3+\sigma|}&\bar{\epsilon}^{|2n_3+\sigma|}\\
\end{array}
\right], \nonumber \\
m^\nu_D &\propto &  \left(
  \begin{array}{ccc}
    \bar\epsilon^{|1 \pm n_1|} & \bar\epsilon^{|1 \pm n_2|} & \bar\epsilon^{|1 \pm  n_3|} \\
    \bar\epsilon^{|n_1|} & \bar\epsilon^{|n_2|} & \bar\epsilon^{|n_3|} \\
    \bar\epsilon^{|n_1|} & \bar\epsilon^{|n_2|} & \bar\epsilon^{|n_3|} 
  \end{array}
\right) , m_{eff }\propto  \left( 
\begin{array}{ccc}
{\bar{\epsilon}^2} & {\bar{\epsilon}} & {\bar{\epsilon}}
\\ 
{\bar{\epsilon}} & 1 & 1 \\
\bar{\epsilon} & 1 & 1
\end{array}
\right).
\label{EX2}
\end{eqnarray}
Appropriate choices of the unspecified U(1) charges, such as
$n_1=2, n_2=-1, n_3=1, \sigma=-1$
(where the $n_i$ are the U(1) charges of the 
right-handed neutrinos, and
$\sigma$ is the U(1) charge  of the field $\Sigma$)
lead to interesting phenomenology.

We use an indicative choice of coefficients for these SU(5) textures given in~\cite{EGL}, 
which is summarised in Table~\ref{table1} below.

\begin{table}[!h]
\begin{center}
\begin{tabular}{|l||l|}
\hline
& Parameters in an SU(5) model with large $\tan\beta$   \\ \hline \hline
Charged leptons &
$a^e_{12}= 0.6, a^e_{13}= 0.9,
a^e_{22}= 1.2,  a^e_{23} = -0.5e^{i \pi/3}$ \\ &
$ a^e_{31}=0.7, a^e_{32}=0.6, a^e_{33} = 0.4$ \\
\hline
$m^\nu_D$ &
$a^\nu_{12} =1.3, a^\nu_{21} =-1.3,  
 a^\nu_{22}= 0.7 $ \\
&  $ a^\nu_{23} = 1.8 e^{i \pi/5}, 
a^\nu_{32}= 0.7, a^\nu_{33}= 0.5 $ \\
\hline
$M_{RR}$ &
$a^N_{22} = 1, a^N_{33} =1.8$ \\
\hline 
\end{tabular}
\caption{\small \it   Choice of coefficients that reproduce the fermion data for an SU(5) model 
with large $\tan\beta$. Coefficients not listed in the table are set to unity.}
\label{table1}
\end{center}
\end{table}

\subsection{Renormalization above $M_{GUT}$}
\label{aboveMGUT}

One must also keep in mind that the GUT scale 
may lie significantly below the
scale $M_{Grav}$ at which gravitational effects can no longer be 
neglected~\footnote{$M_{Grav}$ might be identified with either the Planck mass
$M_P = 1.2 \times 10^{19}$~GeV or some lower string unification scale
$M_{string} \sim 10^{18}$~GeV.}. In this case, the 
renormalization of couplings
at scales between $M_{Grav}$ and $M_{GUT}$ may induce significant
flavour-violating effects, particularly in the right-handed slepton mixing,
which is suppressed in minimal schemes such as that described in the previous subsection.
The simplest such example is provided by the minimal supersymmetric
SU(5) GUT, where the superpotential contains terms 
of the form $\bar{E}\bar{U}\bar{H}$ (with
$\bar{H}$ being a colour-triplet Higgs field that is expected to have a 
mass $\sim M_{GUT}$). This gives rise to one-loop diagrams that
renormalize the right-handed slepton masses.
In the leading-logarithmic approximation, these 
corrections take the form~\cite{HisNo}:
\begin{eqnarray}
(m^2_{\tilde{e}})_{ij} 
&\simeq& - \frac{3}{8\pi^2}
  \lambda_{u_3}^2 V_{U}^{3i} V^{\ast 3j}_{U} 
 (3 m_0^2 +a_0^2)  \log  \frac{M_{\rm grav}}{M_{\rm GUT}} ,
 \label{VU}
\end{eqnarray}
for $i \ne j$, where $V_{U}$ denotes the 
mixing matrix in the corresponding couplings in the basis where
the $u$-quark and charged-lepton masses are diagonal. 
Similarly, the complete leading-logarithmic renormalization 
of the $A_e$ terms is given by 
\begin{eqnarray}
A_e^{ij}&\simeq&
-\frac{3}{8\pi^2} a_0 \left( 
  \lambda_{e_i} V_{D}^{\ast 3i} V_{D}^{3j} \lambda_{\nu_3}^2 
       \log \frac{M_{\rm grav}}{M_{\nu_3}}  
+ \lambda_{e_i} V_{D}^{\ast 2i} V_{D}^{2j} \lambda_{\nu_2}^2 
       \log \frac{M_{\rm grav}}{M_{\nu_2}}  
\right. \nonumber \\
&&
\phantom{-\frac{3}{8\pi^2} a_0 }
\left. + 3 \lambda_{e_j} V_{U}^{\ast 3j} V_{U}^{3i} \lambda_{u_3}^2 
\log \frac{M_{\rm grav}}{M_{\rm GUT}} \right) .
\label{AVU}
\end{eqnarray}

For the structure of the mixing matrices $V_{U,D}$, one has to 
go to a specific GUT model. Within the minimal SU(5) GUT, the $d$-quark
mass matrix is the transpose of the charged-lepton mass matrix, $V_D$ is
simply the unit matrix, and $V_U$ is related to the
familiar CKM matrix. This mechanism therefore provides relatively little mixing.

\subsection{Non-Minimal GUT Effects}
\label{nonminGUT}

In the minimal SU(5) case, the renormalization would be too small to generate observable mixing for
universal initial conditions. However, minimal SU(5) also predicts the unsuccessful
relations $m_s = m_\mu$ and $m_d = m_e$. These  can be modified by 
non-renormalizable terms in the effective superpotential, such as the fourth-order
term ${\tt{10}}-{\tt{24}}-{\tt{\bar 5}}-\bar{H}$, which make
different contributions to the $d$-quark and charged-lepton mass matrices~\cite{ELR},
such as:
\[
\lambda ({\mathbf {10}}-{\mathbf {\bar 5}}-\bar{H}) + \lambda^{\prime} 
(\bar{H}-{\mathbf {10}}-{\mathbf {24}}-{\mathbf {\bar 5}}) 
\rightarrow 
\lambda \bar{v} (d d^c + e^c e) +
\lambda^\prime \bar{v} V (2 d d^c -3 e^c e) .
\]
In this case, in the basis where $m_d$ is diagonal, one has
$m_e = m_d^D - 5 \lambda^{\prime} \bar{v} V$,
where the matrix of couplings $\lambda^{\prime}$ is non-diagonal, in general.
The diagonalization of $m_e^D = V_{eR} m_e V_{eL}^{+}$ then gives
$m_e^D = V_{eR} (m_d^D - 5 \lambda^{\prime} \bar{v} V) V_{eL}^{+}$.
Similarly ~\cite{his2}, 
the colour-triplet-induced $e^c u^c$ mixing may receive large 
corrections for the first two generations. Parametrizing the non-renormalizable
correction to this mixing by $V_{uR},$ the RGE-induced right-slepton mixing
is not given by $V_{CKM}$ as in the minimal model, but by the product 
$V_R=V_{CKM} V^{+}_{uR}$, 
where $V_{uR}$ is  a
potential extra source of right-handed charged-lepton mixing,
that can be significantly larger than what is expected in 
the minimal scheme~\cite{ELR}~\footnote{These 
non-renormalizable corrections also change
the forms of the fermion mass matrices,
and hence the predictions of this type of flavour-texture model within
minimal SU(5). Thus, these corrections would also affect the
renormalization between the GUT and heavy-neutrino mass scales.
Such effects would have supplementary effects on the the left-handed slepton mixing,
but the detailed study of those effects goes beyond the scope of this paper.}.


\section{Discussion of the Supersymmetric Parameter Space}
\label{SUSY-space}

\subsection{First Considerations}

In order to identify representative points in the supersymmetric parameter space
for which detailed simulations are to be run, we first recall a point considered
previously in the literature~\cite{HP}~\footnote{This point is now excluded by the 
LEP bound on the Higgs mass. Nominally, $m_h >114.4$~GeV, but we consider
supersymmetric points with $m_h$ as low as 111~GeV to be acceptable, so as to make
a suitable allowance for uncertainties in the theoretical calculation of $m_h$.}. In this work,
the following CMSSM point has been studied: 
\beq
tan\beta=10,\;\;\;\; m_0=100\rm{GeV}, \;\;\;\; 
M_{1/2}=300\rm{GeV}, \;\;\;\; A_{0}=300\rm{GeV} .
\eeq
This selection of parameters is displayed as point A in Table 2.

\begin{table}[!h]
\begin{center}
\begin{tabular}{| c | c | c | c | c | c | c | c | c |}
\hline
$Point$ & $Model type$ & $m_0$ & $M_{1/2}$ & $\tan\beta$ & $A_0$ & $N_{events}$ & $\sigma_{int}$ & $L_{int}$ \\
\hline
\hline
A & CMSSM & $100$ & $300$ & $10$ & $300$ & 757K & 25.3~pb & 30~fb$^{-1}$ \\
\hline
B & SU(5) & $40$ & $450$ & $35$ & $40$ & 730 K & 2.44~pb & 300~fb$^{-1}$\\ 
\hline
C & CMSSM & $220$ & $500$ & $35$ & $220$ & 536 K & 1.79~pb & 300~fb$^{-1}$ \\
\hline

\hline 
\end{tabular}

\caption{\small \it   Parameters of the two reference points A and B (masses in GeV). 
We also quote the numbers of events simulated, the LHC 
cross sections and the assumed sample luminosities. 
Point A is a CMSSM model with universal soft supersymmetry-breaking terms at the GUT scale.
Point B is a model with universality assumed at a scale $2\cdot 10^{17}$~GeV;  
for comparison with this point, we also present point C,
a set of CMSSM parameters that leads to a similar sparticle
spectrum and satisfies all the cosmological
and phenomenological bounds. In all cases, we work with $\mu>0$.
}

\end{center}
\label{tab:a}
\end{table} 

\begin{table}[!h]
\begin{center}
\begin{tabular}{| c | c | c | c | c | c | c | c | c | c |}
\hline 
$Point$ & $M_{\tilde{g}}$ & $M_{\tilde{u}_L}$ & $M_{\tilde{d}_L}$ & $M_{\tilde{\chi}_2^0}$ & $M_{\tilde{\tau}_1}$ & $M_{\tilde{\chi}_1^0}$ & $M_{\tilde{l}_R}$ & $M_{\tilde{l}_L}$ & $M_{h}$ \\
\hline
\hline
A & $720$ & $664$ & $669$ & $216$ & $150$ & $118$ & $155$ & $232$ & $110$ \\
\hline
B & $1095$ & $1025$ & $1024$ & $366$ & $207$ & $194$ & $286$ & $371$ & $117$\\ 
\hline
C & $1154$ & $1074$ & $1078$ & $388$ & $219$ & 
$206$ & $290$ & $405$ & $116$ \\
\hline
\end{tabular}

\caption{\small \it   Relevant sparticle masses (in GeV) for the 
reference points defined in Table 2.}

\end{center}
\label{tab:b}
\end{table} 

\begin{figure}[!h]
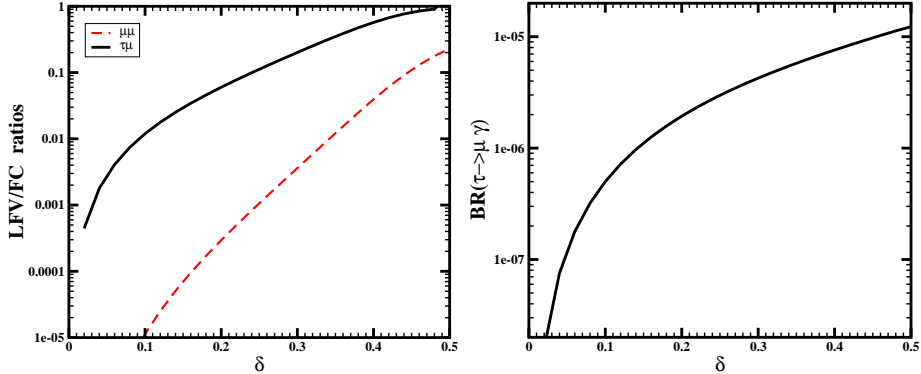

\begin{center}
  \includegraphics[height=.22\textheight]{lfv/hin_fig1.eps}
\includegraphics[height=.22\textheight]{lfv/tmg_hin.eps} 
\end{center}
\caption{\small \it   In the left panel, flavor-conserving and -violating dilepton branching ratios 
are calculated for point A, for comparison with Fig.~1 of~\protect\cite{HP}. The corresponding
expectations for $\tau \to \mu \gamma$ decay are shown in the right panel as a 
function of $\delta$. }
\label{fig:HP}
\end{figure}

We consider first the effects of lavor-mixing entries on the slepton mass matrices that are 
introduced to mimic the non-diagonal terms induced in the $M_{LL}^2$ sector 
by a generic `seesaw' machanism:

\beq
\left(M_{LL}^2\right)_{23}=\delta \cdot \left(M_{LL}^2\right)_{22} ,
\eeq
including all the contributing on-shell sfermion exchange diagrams, 
as given in~\cite{BHH}:
\beq
BR(\chi_2\rightarrow\chi \tau^\pm \mu^\mp)=\sum_{i=1}^3 \left[
BR(\chi_2\rightarrow\tilde{l}_i \mu)BR(\tilde{l}_i\rightarrow\tau \chi) + 
BR(\chi_2\rightarrow\tilde{l}_i \tau)BR(\tilde{l}_i\rightarrow\mu \chi)
\right] .
\eeq
We present in the left panel of Fig.~\ref{fig:HP} the ratio of the flavor-violating decay 
width $\Gamma(\chi_2\rightarrow\chi+ \tau^\pm+ \mu^\mp)$ 
to the CMSSM flavor-conserving decay width 
$\Gamma(\chi_2\rightarrow\chi+ \tau^\pm+ \tau^\mp)$ for $\delta=0$. 
We find similar values to those 
in~\cite{HP}, where only the dominant 
$\tilde{\tau}_1$ exchange diagram was included.

As shown later, in order to have significant LFV signals, we need 
$\Gamma(\chi_2\rightarrow\chi+ \tau^\pm+ \mu^\mp)/
\Gamma(\chi_2\rightarrow\chi+ \tau^\pm+ \tau^\mp) \sim 0.1$. 
We see in the left panel of Fig.~\ref{fig:HP} that this ratio corresponds to 
$\delta\sim 0.25$. However, a potential problem for such points arises from
BR($\tau \rightarrow \mu \gamma$), which tends to become too large. This is shown in
the right panel of Fig.~\ref{fig:HP}, in which 
typical values are much larger than the estimated value of
BR($\tau\rightarrow \mu \gamma$)$\sim 10^{-9}$ in~\cite{HP}. We recall that the current 
experimental upper bound is $<6.8 \cdot 10^{-8}$~\cite{pdg}.

When a similar choice of CMSSM parameters is considered in the framework of 
an SU(5) model with seesaw neutrinos like the one described in 
Sections 3 and 4 we find that by setting $M_R=3 \cdot 10^{14}$~GeV we obtain a value 
of $\delta\sim 0.05$ leading to a prediction 
BR($\tau\rightarrow \mu \gamma$) $\sim 3 \cdot 10^{-8}$. 
This would be acceptable for $\tau\rightarrow \mu \gamma$, but would not lead to
observable $\chi_2\rightarrow\chi+ \tau^\pm+ \mu^\mp$ decay.
Due to the strong bound imposed by $\tau\rightarrow \mu \gamma$, 
it is very difficult to obtain reasonable values 
of  $\delta$ using  only the LL mixing found in seesaw models.

However, mixing in the RR sector can enhance the decays so that 
they might be detectable at the LHC without a large 
increase in BR($\tau\rightarrow \mu \gamma$), since RR mixing enters only in 
the subdominant one-loop neutralino-exchange diagram for this process. 
It is considerably smaller than the dominant
chargino-exchange diagram for this process, which is sensitive to LL mixing.
However, as we have already discussed, significant mixing in the RR sector of the slepton 
mass matrix cannot be obtained in minimal SU(5) with the conventional seesaw
mechanism, but may be obtained in more general GUT 
scenarios~\footnote{Large mixing in the
RR sector was also found to be necessary for flavour-violating effects to be
observable via non-universality in leptonic kaon decays~\cite{ELR}.}

\begin{figure}[!h]
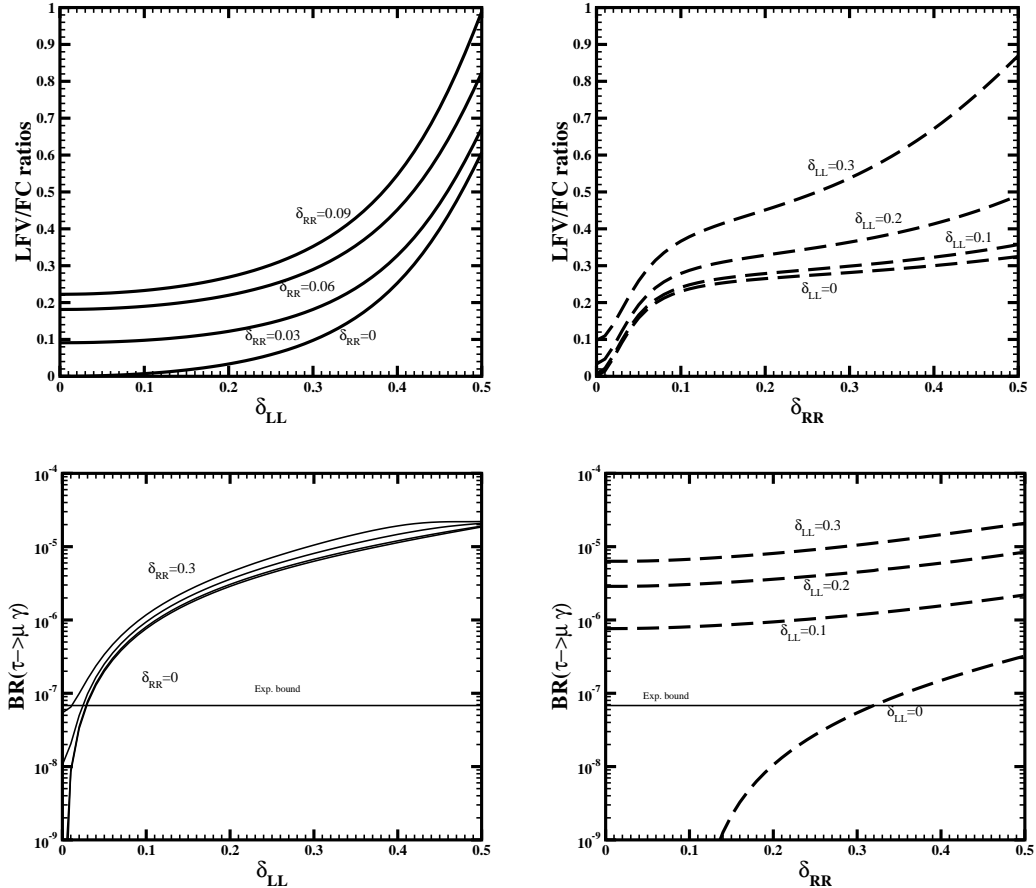

\begin{center}
  \includegraphics[height=.25\textheight]{lfv/ratdel_LL.eps}
\hspace{.5 true cm}
  \includegraphics[height=.25\textheight]{lfv/ratdel_RR.eps} 

\vspace*{0.5cm}
\includegraphics[height=.25\textheight]{lfv/brdel_LL.eps}
\hspace{.5 true cm}
  \includegraphics[height=.25\textheight]{lfv/brdel_RR.eps} 
\end{center}
\caption{\small \it  Branching ratios for point A for $\chi_2\rightarrow\chi+ \tau^\pm+ \mu^\mp$ decay
(upper plots) and $\tau\rightarrow \mu \gamma$ (lower plots) as functions of $\delta_{LL}$
(left) and $\delta_{RR}$ (right) for certain discrete choices of $\delta_{RR}$ (left) and
$\delta_{LL}$ (right), for the same CMSSM inputs as in Fig.~1.}
\label{ratdel}
\end{figure}

In order to see how a branching ratio for $\chi_2\rightarrow\chi+ \tau^\pm+ \mu^\mp$
decay that could be observable at the LHC might be induced by non-diagonal 
entries in the $M_{RR}$ sector of the charged-slepton mass matrix, without violating the 
$\tau\rightarrow \mu \gamma$ bound, we parametrize as follows
the LFV entries in the charged-slepton mass matrices:
\bea
\left(M_{LL}^2\right)_{23}&=&\delta_{LL} \cdot \left(M_{LL}^2\right)_{22}, \nonumber \\
\left(M_{RR}^2\right)_{23}&=&\delta_{RR} \cdot \left(M_{RR}^2\right)_{22}.
\ea
We illustrate in Fig.~\ref{ratdel} the different dependences of the branching ratios for
$\chi_2\rightarrow\chi+ \tau^\pm+ \mu^\mp$ and $\tau \to \mu \gamma$ decays on the magnitudes
of the off-diagonal LL and RR LFV mixing entries.

The upper plots in Fig.~\ref{ratdel} display the dependences of the
$\chi_2\rightarrow\chi+ \tau^\pm+ \mu^\mp$ decays of interest, and the lower
plots display the dependences of the branching ratio for
$\tau\rightarrow \mu \gamma$ decay. The left plots show the dependences on
$\delta_{LL}$ for certain discrete choices of $\delta_{RR}$, and the roles of
LL and RR mixing are reversed in the right plots. We see again that if
RR mixing were negligible 
a 10\% ratio of the LFV $\chi_2\rightarrow\chi+ \tau^\pm+ \mu^\mp$ decays
relative to the flavour-conserving ones would require large LL mixing
with $ \delta_{LL}\sim 0.25$,  which would inevitably imply a violation of the 
$BR(\tau\rightarrow \mu \gamma)$ bound. On the other hand, 
we see that a value of   
$ \delta_{RR}\sim 0.02-0.03$ would lead to a similar ratio for
$\chi_2\rightarrow\chi+ \tau^\pm+ \mu^\mp$ decay, whatever the value of $\delta_{LL}$,
whilst the LFV radiative tau decay could remain within the acceptable
experimental range if $\delta_{LL}$ were below about 0.03. 
We also observe 
that  the size of $ \delta_{RR}$ required to obtain the desired ratios 
increases as $\tan\beta$ increases, requiring physics beyond
the minimal SU(5) model, as discussed in 
Section~\ref{nonminGUT}.

\subsection{Phenomenological and Cosmological Constraints}

The parameter point A discussed in the previous section predicts a neutralino 
relic density that exceeds the WMAP upper  bound, and also predicts 
$m_h<111$~GeV, below even the more conservative LEP bound on  the 
Higss  mass allowing for theoretical uncertainties. 
In this subsection we specify a model that satisfies these 
phenomenological constraints, and may also incorporate
the non-minimal SU(5) GUT model with a seesaw 
mechanism described earlier.  

We assume universal soft supersymmetry breaking at a scale 
$M_X = 2 \cdot 10^{17}$~GeV$ >M_{GUT}$, in which case we find a 
minimum value of $\tan\beta\sim 31 $ for $\mu>0$, above which  
$\tilde{\tau}-\chi$ coannihilations are sufficient to bring $\Omega_\chi h^2$ below the 
upper WMAP bound~\cite{mambrini}. Furthermore, we 
find that a minimum value $\tan\beta\sim 35$ is needed to ensure $m_h>114$~GeV,
though we recall that we assign a theoretical uncertainty of $\sim 3$~GeV to
the theoretical calculation of $m_h$.
The GUT Yukawa coupling relation $Y_\tau=Y_b$ cannot 
be achieved for $\mu>0$~\cite{GINS} due to the large supersymmetric threshold 
corrections to $m_b$. However, lepton mixing effects may modify the GUT relation: 
\beq
Y_b(GUT)=Y_\tau(GUT)(1-x) ,
\label{xpa}
\eeq
where $x$ is a parameter that accounts for a sizeable 2-3 generation mixing in the 
charged-lepton Yukawa coupling matrix~\cite{LLR, CELW,pedro}. 

\begin{figure}
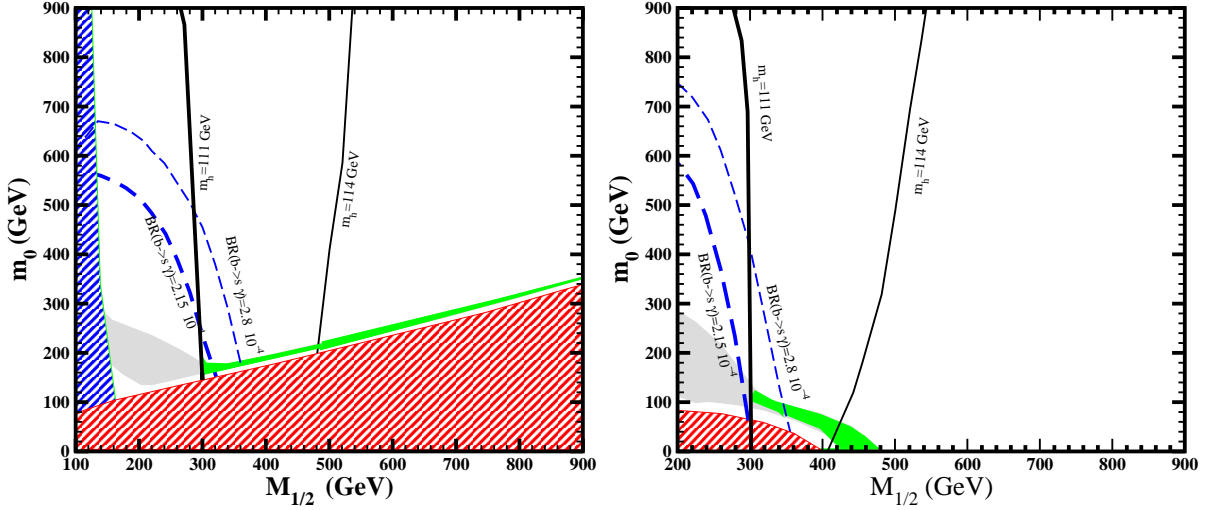

  \includegraphics[height=.3\textheight]{lfv/m0m12_tb35g.eps}
  \includegraphics[height=.3\textheight]{lfv/m0m12_tb35x.eps} 
  \caption{\small \it   Cosmologically-favored areas (green) in 
the $(M_{1/2}, m_0)$ 
plane for $\tan\beta=35$ and $A_0=m_0$, assuming SU(5) unification. In the 
left panel we assume universality at $M_X=M_{GUT}$, whereas in the 
right panel we assume universality at $M_X = 2 \cdot 10^{17}$~GeV. 
The red areas are excluded because $m_\chi> m_{\tilde{\tau}}$. We also display the 
contours for $m_h=111, 114$~GeV (black solid and thin solid) and 
$BR(b\rightarrow s \gamma)\cdot 10^{4}<2.15, 2.85$ (blue dashed and thin dashed). 
}
\label{fig:areas}
\end{figure}

We display in Fig.~\ref{fig:areas} two $(M_{1/2}, m_0)$ planes for the same choice 
$\tan\beta=35$. In the left panel, we assume universality at $M_X=M_{GUT}$, 
whereas in the right panel we show the changes in the allowed parameter
space when $M_X=2 \cdot 10^{17}$~GeV. In the latter case, we find a region of parameter 
space in which the WMAP bound on the cold dark matter density is respected simultaneously
with the conservative bound $m_h>114$~GeV. In both cases, we fix 
$m_b(M_Z)=2.92$~GeV, which corresponds to the evolution 
of the  central value of the $\overline{MS}$ value 
$m_b(mb)=4.25$~GeV with $\alpha_s(M_Z)=0.172$. With this choice of $m_b$, we 
find a value $x=0.37$ in (\ref{xpa}).

In order to take into account the cosmological and phenomenological considerations
discussed above, we now study point B in Table~2,
whose sparticle spectrum is tabulated in Table~3.
Point A is taken from~\cite{HP} and is included
so as to facilitate comparisons. 
As we shall see, point B
allows for an observable number of events
with the luminosity expected at the LHC.

\begin{figure}[!h]
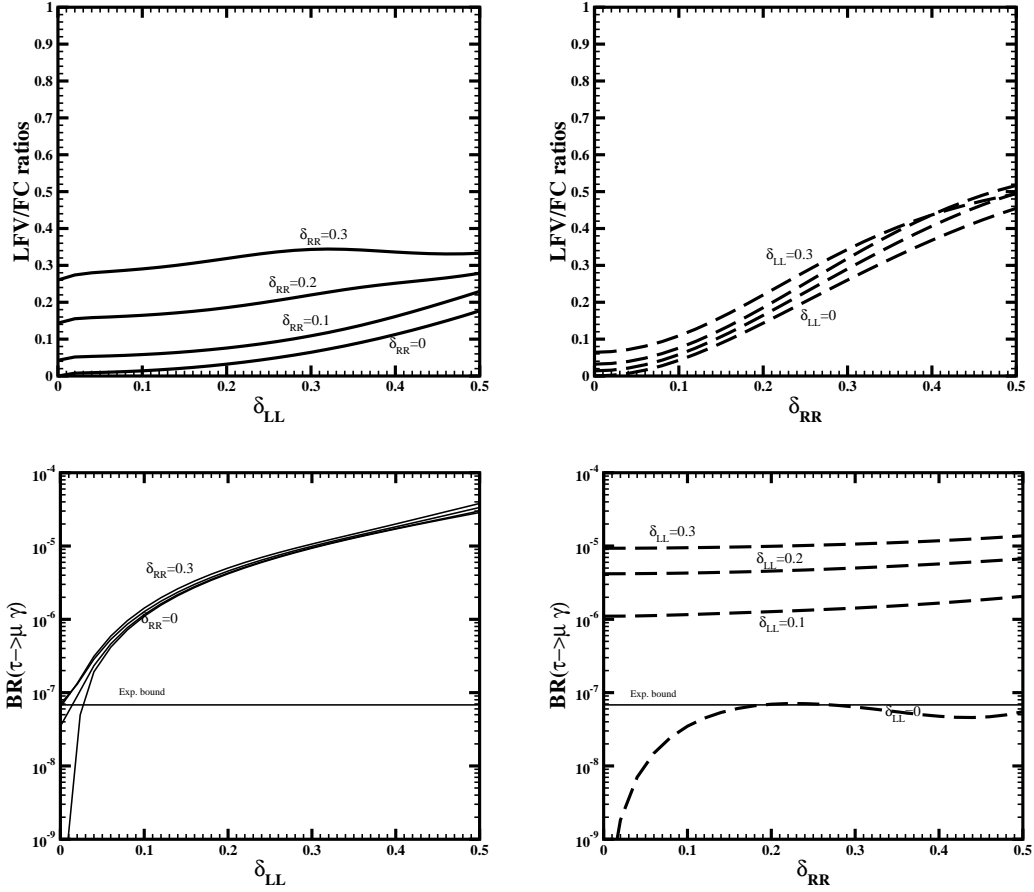

\begin{center}
  \includegraphics[height=.25\textheight]{lfv/ratdel_LLB.eps}
\hspace{.5 true cm}
  \includegraphics[height=.25\textheight]{lfv/ratdel_RRB.eps} 

\vspace*{.5cm}

\includegraphics[height=.25\textheight]{lfv/brdel_LLB.eps}
\hspace{.5 true cm}
  \includegraphics[height=.25\textheight]{lfv/brdel_RRB.eps} 
\end{center}
\caption{\small \it   Branching ratios, as in Fig.~\protect\ref{ratdel},
for the point B described in the text.}
\label{ratdelB}
\end{figure}

We display in Fig.~\ref{ratdelB} for point B
the same sets of branching ratios as were shown in Fig.~\ref{ratdel} for point A.
We see in the top panels that in this case we also need large 
non-diagonal entries in the slepton mass matrix in order to achieve a
branching ratio for $\tilde{\chi}_2\rightarrow \tilde{\chi}_1\tau^\pm\mu^\mp$ that is
of interest for the LHC, e.g., $\delta_{RR} \sim 0.15$ for $\delta_{LL} = 0$
or $\delta_{LL} \sim 0.35$ for $\delta_{RR} = 0$. We also see in the
lower panels that $\tau\rightarrow \mu \gamma$ is again very restrictive on the size of 
$\delta_{LL}$, imposing a maximum value $\sim 0.03$. We see in the bottom-right
panel that $\delta_{RR} \sim 0.15$ is allowed for $\delta_{LL} = 0$~\footnote{If one chose a 
model with (nominally) the lower value $m_h=111$~GeV, one could allow lighter masses 
than for point B, however the limits from BR($\tau\rightarrow \mu \gamma$) 
on $\delta_{RR}$ and  $\delta_{LL}$ would be more severe.}.
The choice $M_X = 2 \cdot 10^{17}$~GeV opens up the possibility of
generating such a value of $\delta_{RR}$ though RGE-induced mixing above $M_{GUT}$.
If we assume a generic pattern of non-universal $M_{RR}^2$ GUT entries 
of the form:
\beq
M_{RR}^2(GUT)= \left(\begin{array}{ccc}
m^2&0&0\\
0&m^2&\epsilon \cdot m_0^2\\
0&\epsilon \cdot m_0^2& m_3^2
\end{array}
\right) ,
\eeq
a value $\epsilon \sim 0.25$ is needed to generate  the value 
$\delta_{RR}\sim 0.15$ at the electroweak scale.

\section{Signals and Backgrounds at the Event Generator Level}

This section is devoted to considerations about the detection 
of LFV in neutralino decays at the LHC, at the
event generator level. The signal of tau flavour violation that we consider is an excess of
dilepton pairs of the type $\mu\tau_h$ over $e\tau_h$ pairs, where $\tau_h$ signifies a hadronic jet 
produced by hadronic tau decay. The hadronic decay channel for the tau lepton provides
a good SUSY signature because: (i) tau leptons decay 65 \% of the time to 
hadrons, and (ii) hadronic jets from tau decays are narrower and have lower particle multiplicity 
than conventional QCD jets, and they can be 
reconstructed from the tracking and calorimeter information.
Although their charges cannot be determined reliably, tau leptons 
decaying hadronically can be detected more clearly than taus decaying to lighter leptons because 
leptons arising from tau decays have two neutrinos, and so are more difficult to reconstruct than
leptonic tau decays and separate from background processes.

We present, analyse and compare results 
from two supersymmetric models with different sparticle spectra, one formulated within the CMSSM and another incorporating additional GUT-inspired physics, 
as summarized in Tables 2 and 3. The spectra for these points
were calculated using {\tt ISAJET~7.78}~\cite{ISAJET} and then interfaced into 
{\tt PYTHIA~6.418}~\cite{Pytia}, which was used to generate the cross sections, the ensuing QCD
parton showers and the hadronic tau decays.

\subsection{General Event Cuts}

There are some general and distinctive features of R-conserving sparticle production at the LHC.
First, the dominant production mechanism is expected to be the pair production of
massive squarks and gluinos, which subsequently decay into hadrons, resulting in several hard jets.
On the other hand, the neutralino LSP escapes detection and thus we expect large missing transverse 
energy (MET) in SUSY events. These features suggest using the following 
definition for the effective mass of  the process:
$M_{eff} \equiv \slsh{\!E}_t + p_{t,1} + p_{t,2} + p_{t,3} + p_{t,4}$,
where $p_{t,i}$ stand for 
the transverse momenta of the four hardest jets. 
The distribution in $M_{eff}$ would exhibit a peak 
around the scales of the squark and gluino masses.

Keeping these features in mind, we apply the following 
cuts on the events generated by 
{\tt PYTHIA}, for both the SUSY signal and the Standard Model background. 

\begin{eqnarray}\label{eq-cuts}
& (i) & N_{jets} \geq 4 \ \mbox{ with} \ p_{T,1} > 100 
\mbox{ GeV and} \ p_{T,2,3,4} > 50 \mbox{ GeV},  \nonumber \\ 
&(ii) & \slsh{E}_t > 0.2 M_{eff} \mbox{  and  } \slsh{E_t} > \slsh{E_t}^{min}, \nonumber  \\ 
&(iii)& M_{eff} \equiv \slsh{E}_t + p_{T,1} + p_{T,2} + 
p_{T,3} + p_{T,3} + p_{T,4} > M_{eff}^{min} \mbox{ GeV}; \nonumber 
\end{eqnarray}
The values of $\slsh{\!E_t}^{min}$ and $M_{eff}^{min}$  were 
fixed independently for each set of 
SUSY parameters, in order to optimize the suppression of the Standard Model background
in each case.

\subsection{Calorimeter and Object Definition}

Only calorimetric 
smearing and segmentation have been incorporated, since 
a full detector simulation is out of 
the scope of the present work. A uniform segmentation 
$\Delta\phi =\Delta\eta=0.1$ is assumed and the energy smearing is done 
according to the following detector parametrization:

\begin{eqnarray}
{\rm ECAL} &\sim& 10\%/\sqrt{E} + 1\%, \mid \eta\mid<3 ,\nonumber\\
{\rm HCAL} &\sim& 50\%/\sqrt{E} + 3\%, \mid\eta\mid<3  ,\nonumber\\
{\rm FCAL} &\sim& 100\%/\sqrt{E} + 7\%, \mid\eta\mid>3 .\nonumber
\end{eqnarray}
We use the following object definitions:
\begin{itemize}
\item Jets are identified with the help of the {\tt PYCELL} subroutine of {\tt PYTHIA}, 
with $R=0.4$ used for the jet cone size, due to the high jet multiplicity 
in SUSY cascade events. We take cells with more than 1~GeV  
as possible jet triggers and require a sum of $\slsh{\!E}_T>10 GeV$ 
in the calorimeter  for the jet to be finally accepted.
\item The total missing transverse energy, $E_T$, is defined as the 
vector sum of the deposits in the calorimeter cells defined above.
\item Leptons are required to be central in rapidity, with
$\mid\! \eta \!\mid<2.5 $, and we use the following isolation demand: no more 
than 10~GeV of transverse energy should be present in a cone size 
of $R=0.2$ around the lepton direction.
\item Hadronic tau decays were selected by requiring 
$p_T > 20$ GeV and $\mid\! \eta \!\mid<2.5$, and a `matching' jet with 
$p_{T}^{\tau} > 0.8 \ p_{T}^{jet}$ and $\mid\!\eta\!\mid<2.5$. 
The jet was tagged as a tau in the two following cases:
(i) if the jet-tau distance is no 
more than $R=0.4$ in the $\eta-\phi$ plane, or
(ii) if the jet-tau distance $R > 0.4$, the jet could also be accepted as a tau,
with a probability
\begin{equation}
P = 1 - \left(0.971p_T^{3/2}-49 \right)^{\frac{5}{3}(1-\epsilon_{\tau})} \nonumber
\end{equation}
where $\epsilon_{\tau}=0.7$ is the selected efficiency.
\item Since {\tt ISAJET} does not allow for LFV decays,
flavor-violating dileptons ($\tau^{\pm}\mu^{\mp}$) 
were simulated by counting events with two taus, 
with at least one of them decaying hadronically; the second 
tau was then tagged as a muon with a probability equal to the 
assumed LFV branching ratio (10 $\%$).
\end{itemize}

\subsection{Standard Model Background}

Standard Model processes that could in principle
contribute to the background for the signature we study here are 
$t\bar{t}$, $WW$, QCD jets, $Z$ jets and $W$ jets. 
The dominant backgrounds are expected to be $t\bar{t}$ and $WW$,
and only these are considered here. In practice, no event coming from the other processes would
satisfy the above-mentioned kinematical requirements when simulated using {\tt PYTHIA}, as used here,
and a more extensive study of Standard Model backgrounds is out of the scope of 
this paper. 
Such a study would require the use of specialized codes for the evaluation of NLO cross sections 
and of the exact matrix elements for the parton showers and hadronic decays. 
The background samples and cross sections used in this work 
are summarized in Table~\ref{count-table1}~\footnote{Because
the simulations are quite time-consuming, rather
than generate the total number of events corresponding to the reference luminosity,
we have rescaled the total number of events by
an overall factor which, in the case of 
100 fb$^{-1}$, is roughly 2.5 times that for 10 fb$^{-1}$.}.

\begin{table}[!h]
\begin{center}
\begin{tabular}{| c | c | c | c |}
\hline
\hline
Process & 10~fb$^{-1}$ & 100~fb$^{-1}$ & $\sigma$ \\
\hline
\hline
$t\bar{t}$ & 250K & 1M & 486~pb$^{-1}$ \\
\hline
$WW$ & 250K & 1M & 70~pb$^{-1}$ \\
\hline
\end{tabular}
\caption{\small \it   Standard Model background samples for reference 
luminosities $10 \ fb^{-1}$ and $100 \ fb^{-1}$, together with their LHC cross sections
calculated using {\tt PYTHIA}}. 
\label{count-table1}
\end{center}
\end{table}

\section{Results for LFV at the LHC}
\label{Results}

In this Section, we analyse the sets of parameters described in 
Table~2, presenting the results of our event simulations. We recall that point A has
been studied previously in~\cite{HP}, and it is included here solely as a consistency check. Point B
is based on the SU(5) RGEs in the framework discussed in previous Sections,
demanding  compatibility with phenomenological and cosmological bounds.
We also recall that point C of Table 2 yields a low-energy spectrum 
and results that are very
similar to those of point B.

Plots corresponding to point A
(usually placed at the tops of the figure arrays), show 
the numbers of events normalized
to a reference luminosity of 10~fb$^{-1}$. 
Plots corresponding to point B
(usually placed at the bottoms of the figure arrays) are for an integrated luminosity of 
100~fb$^{-1}$.

\subsection{Choices of Cuts}
\label{cut-choices}

\begin{figure}[!h]
\begin{tabular}{ccc}
\hspace*{-0.35 in}
\epsfig{file=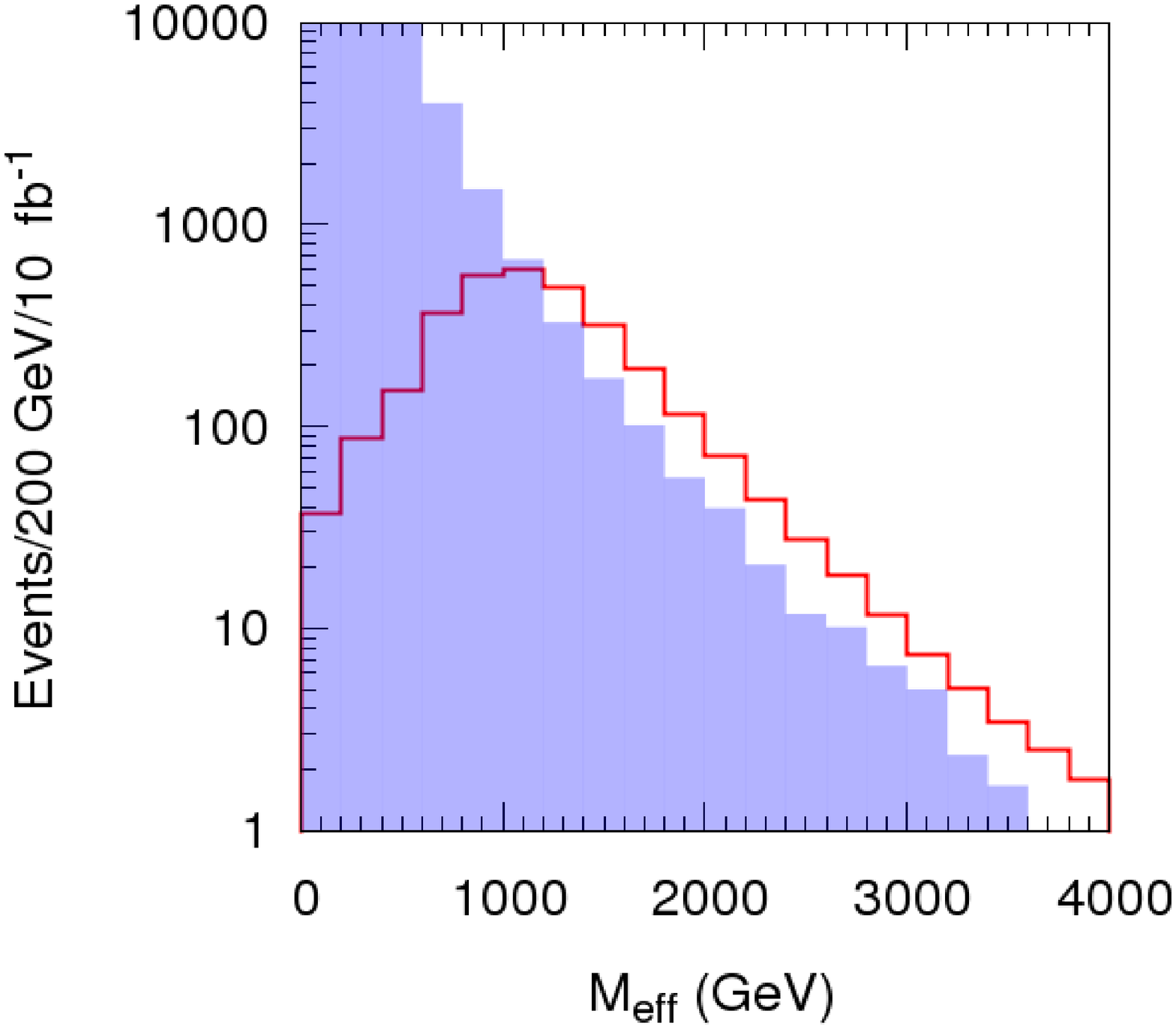,height=2.2 in} &
\epsfig{file=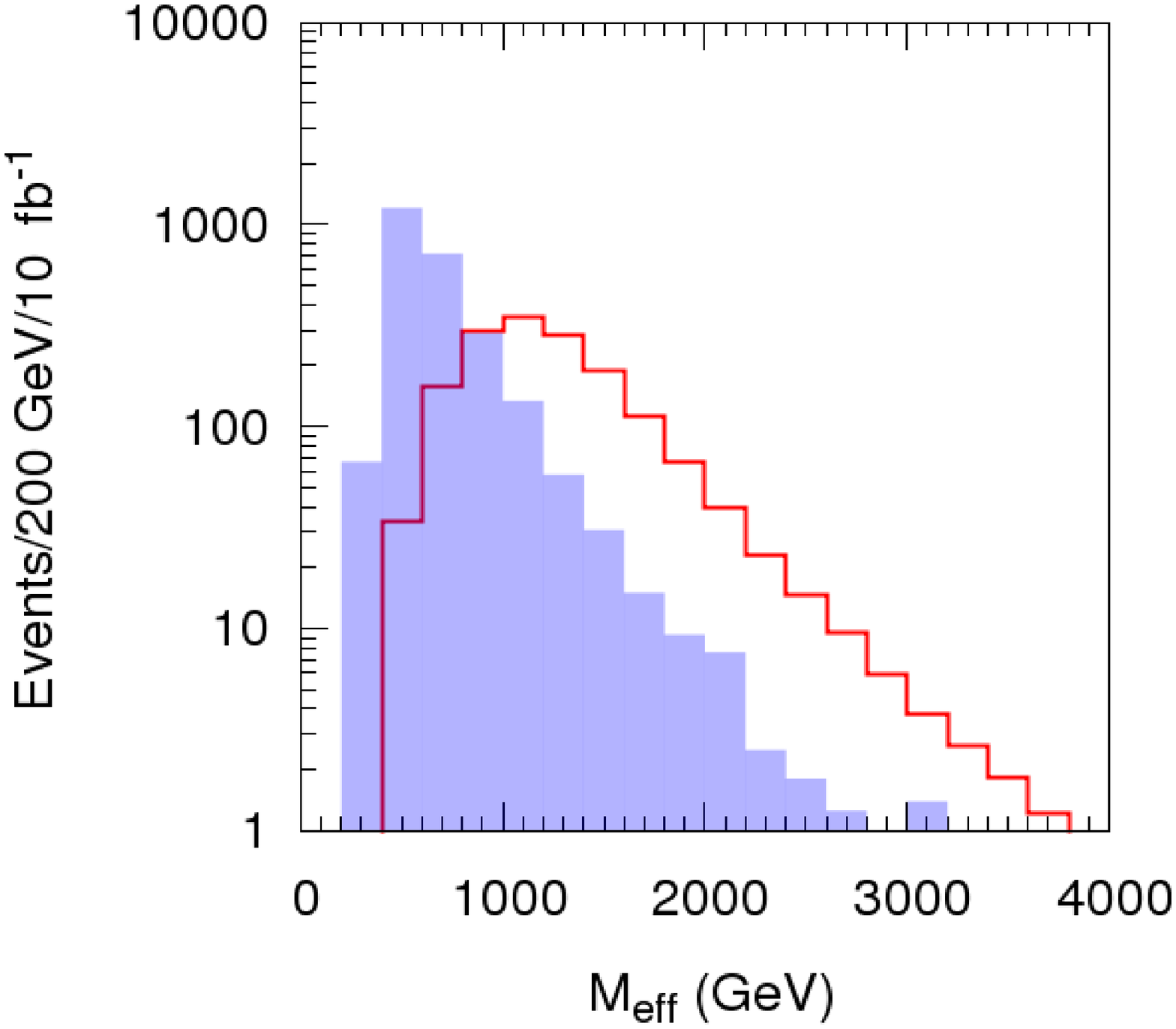,height=2.2 in} &
\epsfig{file=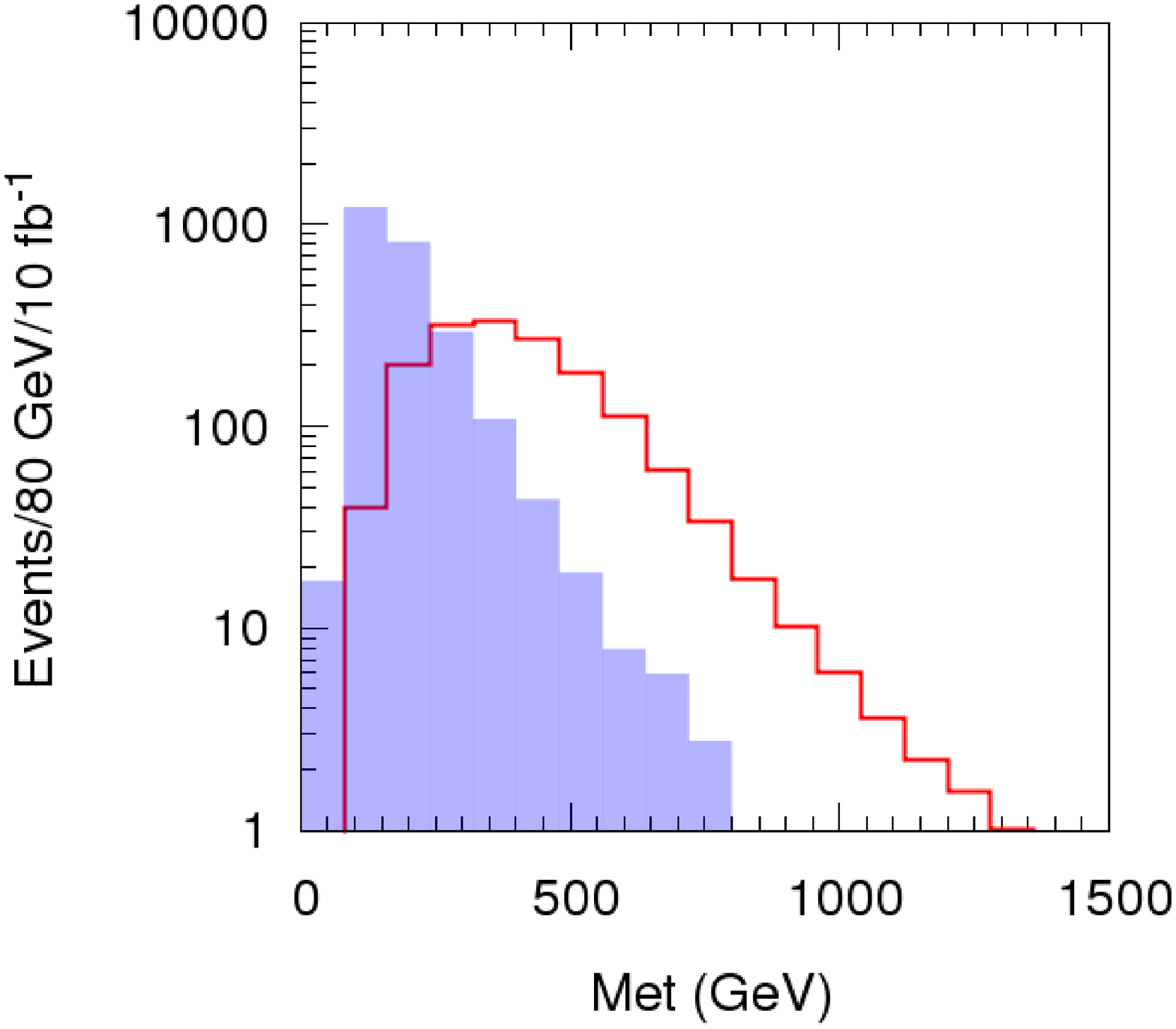,height=2.2 in} 
\\
\hspace*{-0.35 in}
\epsfig{file=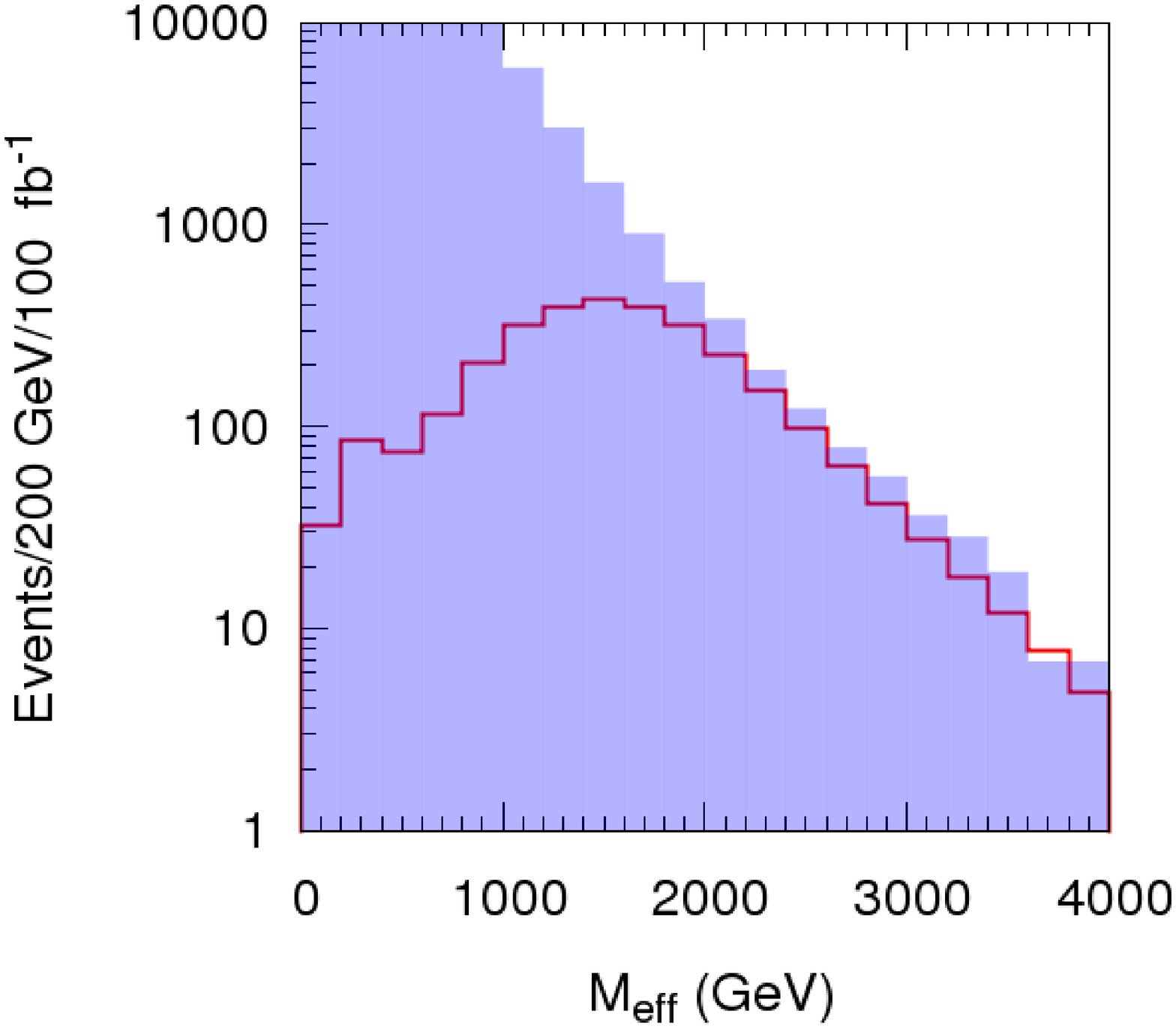,height=2.2 in} &
\epsfig{file=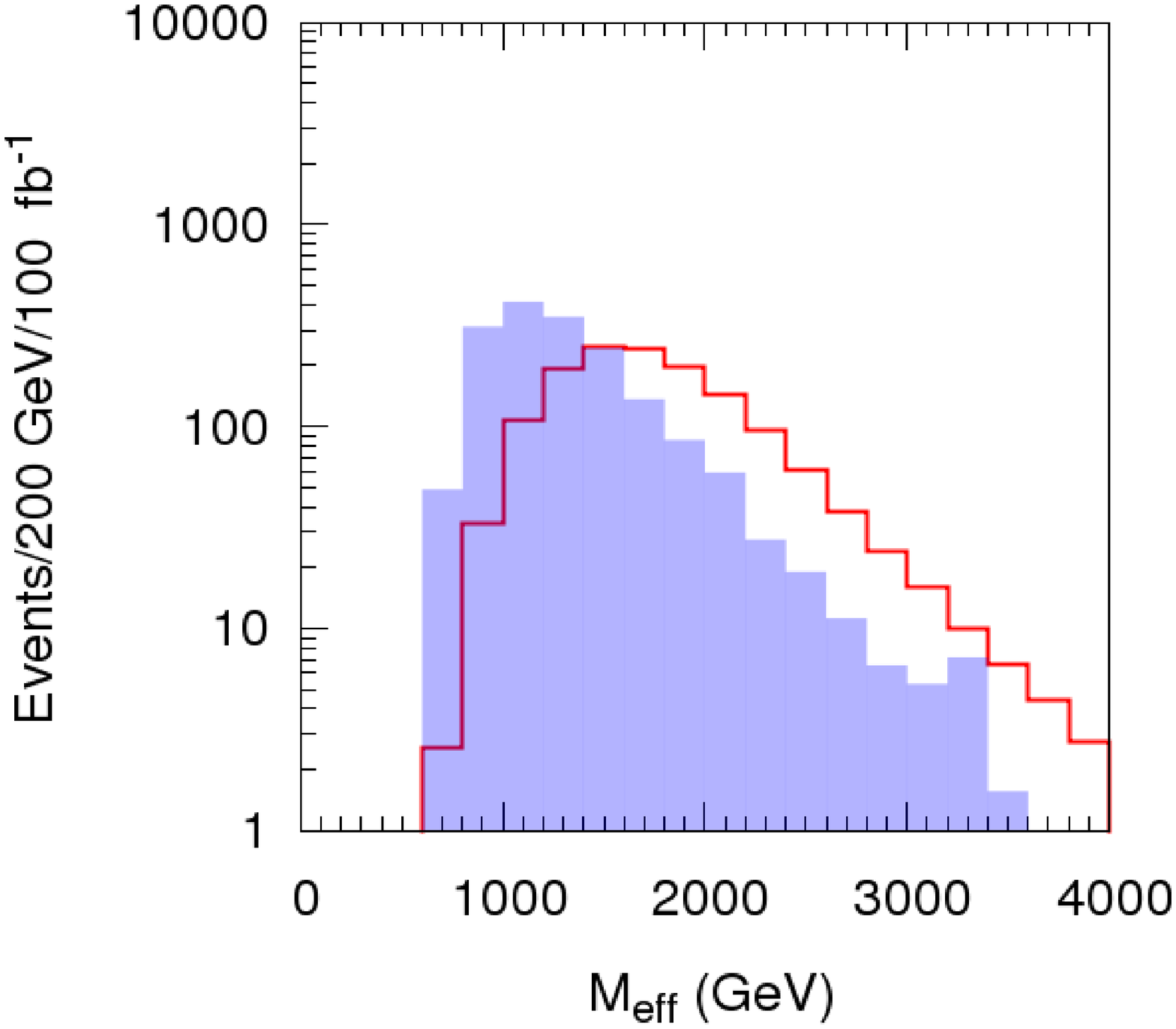,height=2.2 in} &
\epsfig{file=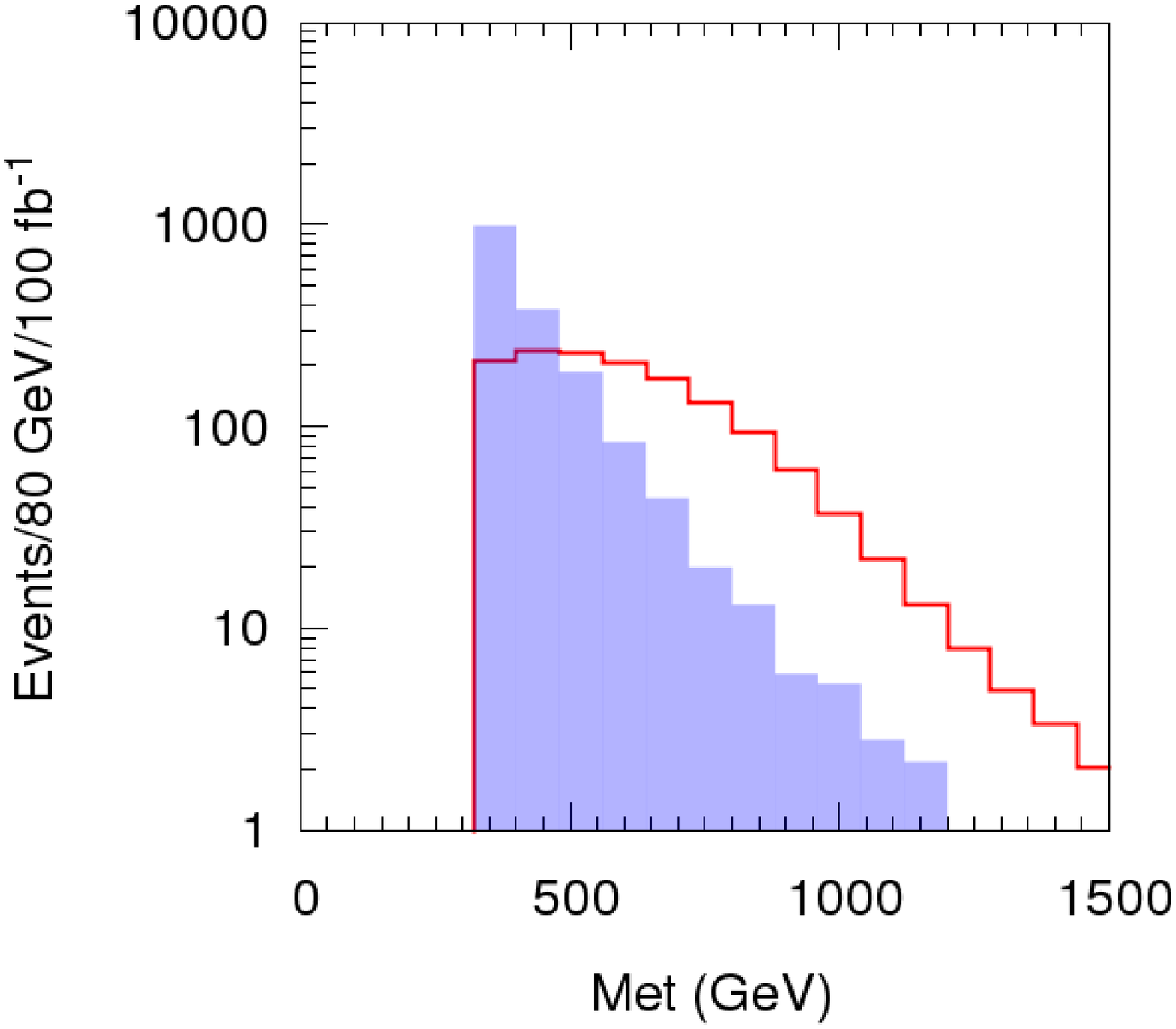,height=2.2 in} \\
(a) & (b) & (c) 
\end{tabular}
\caption{\small \it   $M_{eff}$ distributions for the crude events (plots a) and after  
applying cuts (i) and (ii) (plots b), and the Missing Transverse Energy (MET) 
distributions after applying the same cuts (plots c). The plots on the top correspond 
to simulations for point A, where $\slsh{E}_t^{min}=0$; those on the bottom 
to simulations for point B with $\slsh{E}_t^{min}=320$~GeV. 
The signal distributions are shown as (red) solid lines, and the Standard Model 
backgrounds by the (blue) shaded histograms.}
\label{fig:h1}
\end{figure}

In Fig.~\ref{fig:h1}, we show the $M_{eff}$ and Missing Transverse Energy (MET) distributions 
of events simulated for point A (top) and point B (bottom).
\

 Plots (a) show the distributions calculated including all the events for 
both the Standard Model and SUSY, i.e., without applying any of the cuts in (\ref{eq-cuts}).
We see that the SUSY events exceed the Standard Model ones
at large values of $M_{eff} > 1200$~GeV for point A, whilst  
the distribution for point B shows larger numbers of Standard Model events than SUSY events 
over the entire $M_{eff}$ range. 

In plots (b), after imposing cuts (i) and (ii) (with $\slsh{\!E}_t^{min}$ = 0 
for point A and $\slsh{\!E}_t^{min} = 320$~GeV for point B, the SUSY events 
exceed the SM events at $\approx 800$~GeV and $\approx 1400$~GeV 
for points A and C respectively. We use these values for $M_{eff}^{\rm min}$ in cut (iii) 
in the different cases. 

Plots (c) show the MET distribution after applying the above cuts.
The SUSY event rates exceed those of the Standard Model at energies 
above $\approx 240$~GeV for point A and $\approx 480$~GeV for point B.
Thus, no cut in MET is needed for point A (it would reduce the Standard Model background but also 
the SUSY signal, as can be seen below for the invariant mass distributions, which we denote
by IMD). However, a cut in MET is required for point B, reducing the total number of accepted SUSY 
events for point B relative to point A.

\subsection{Invariant Mass Distributions}

\begin{figure}[!hbt]
\begin{center}
\begin{tabular}{cc}
\epsfig{file=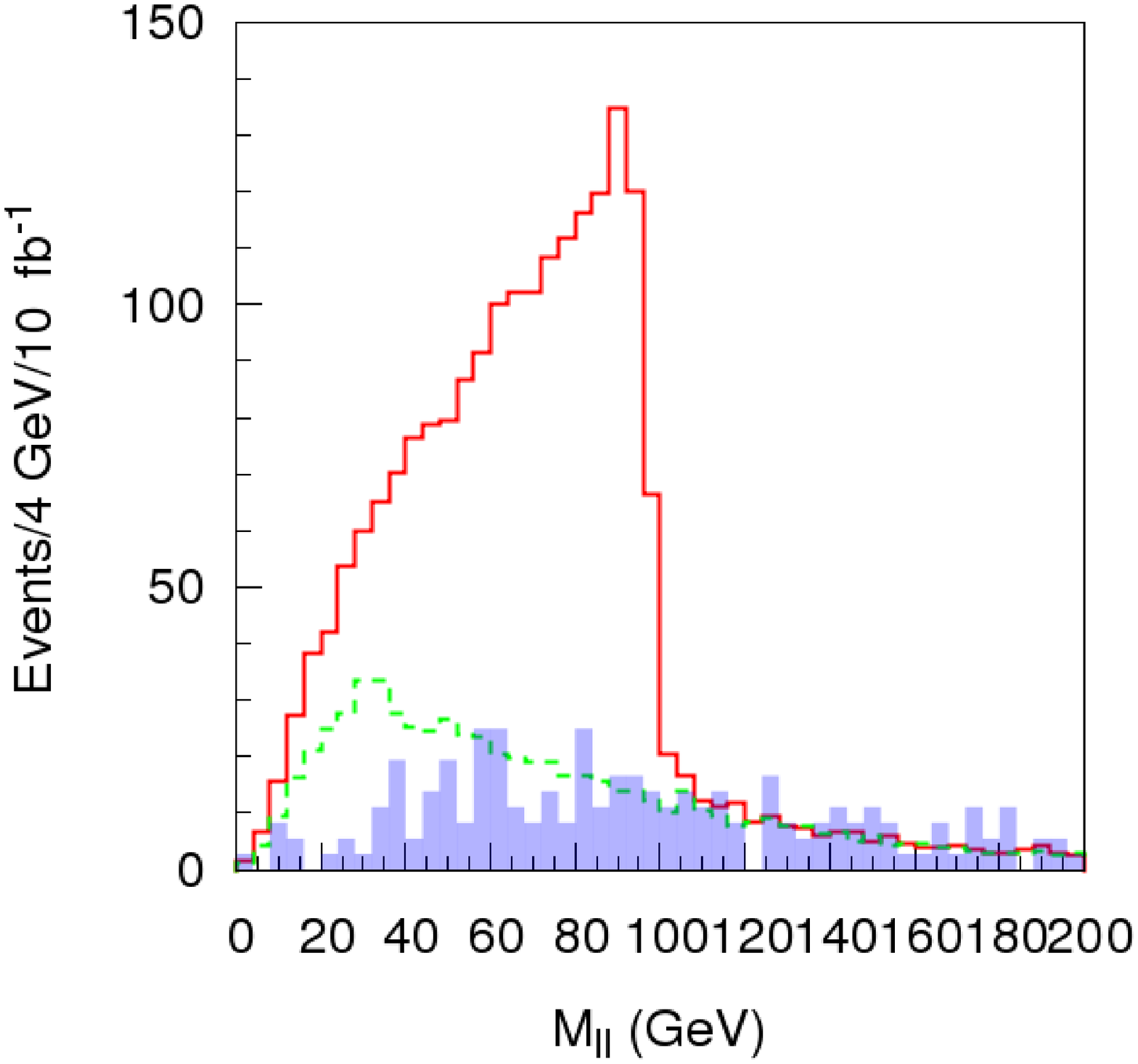,height=2.6in} &
\epsfig{file=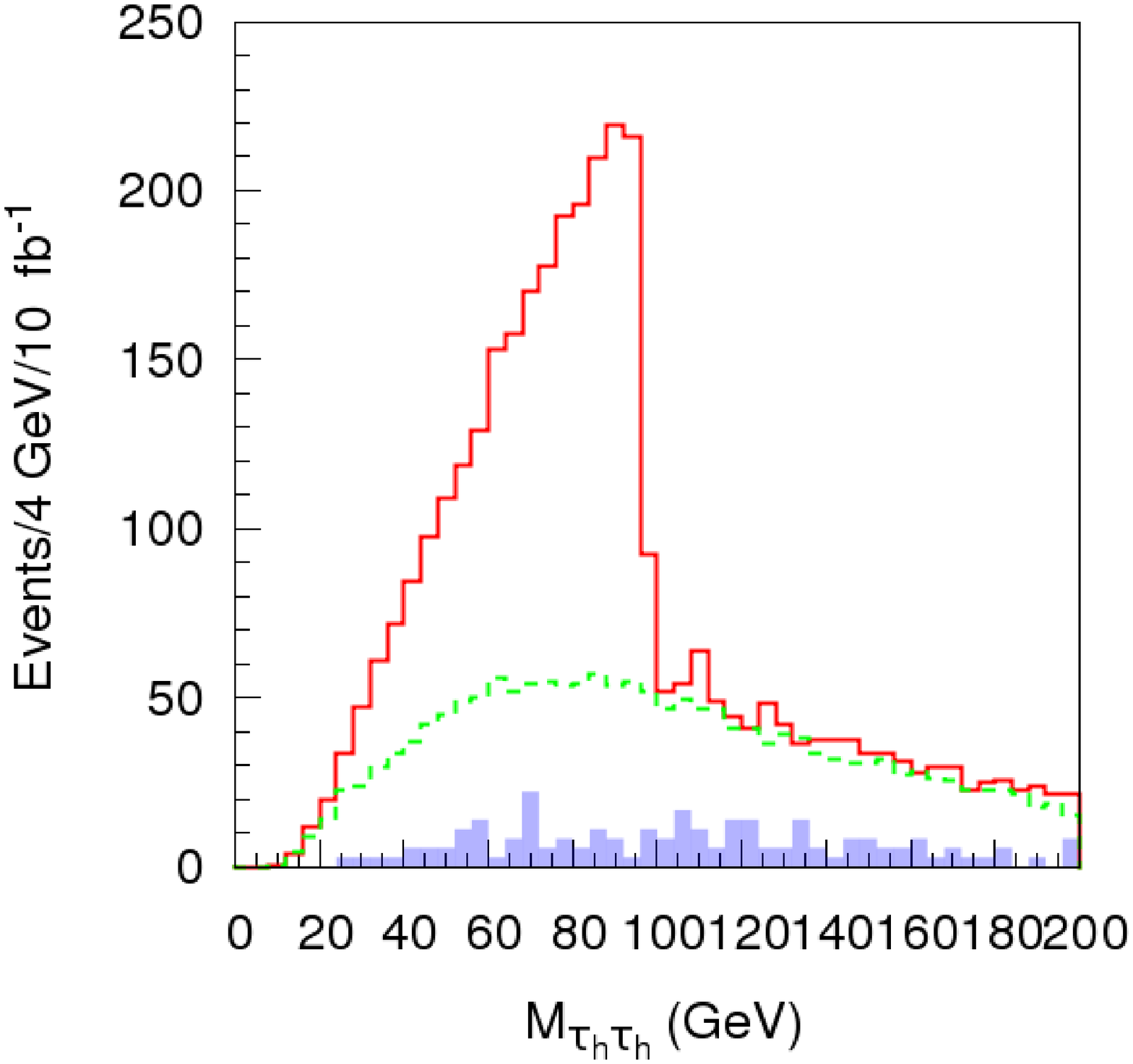,height=2.6in} \\ 
\epsfig{file=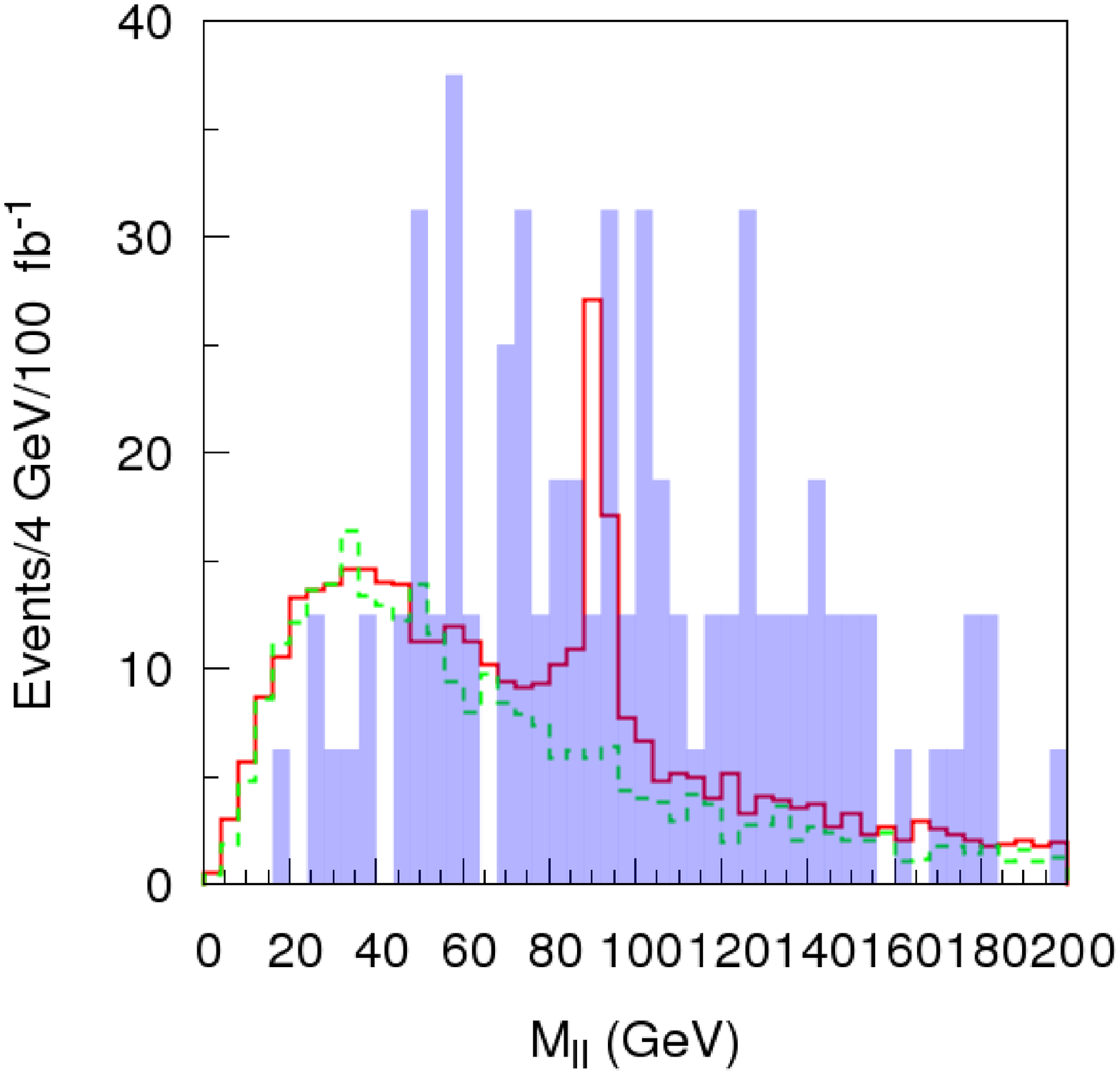,height=2.6in} &
\epsfig{file=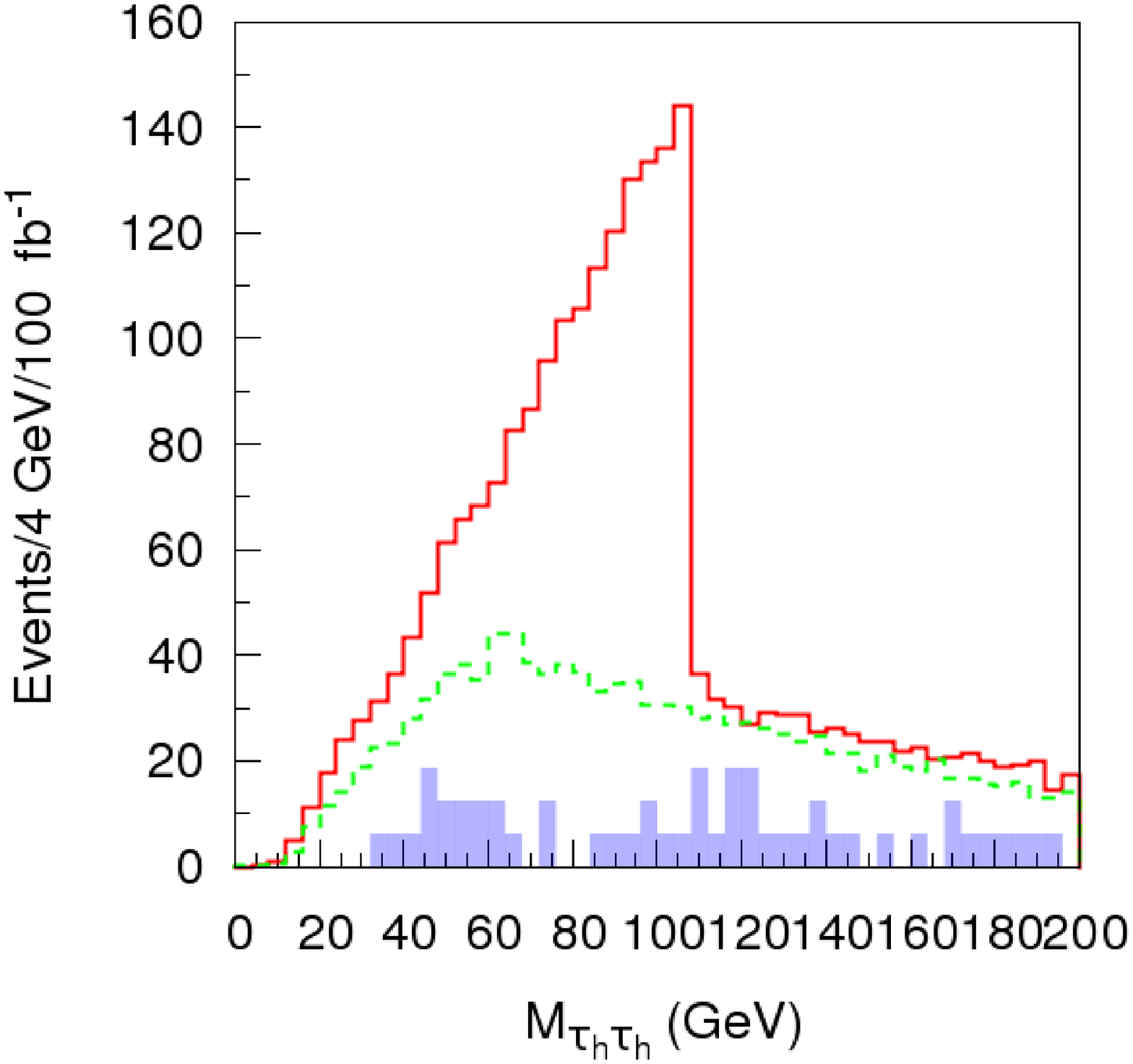,height=2.6in} 
\end{tabular}
\end{center}
\caption{\small \it   (Left) Dilepton invariant mass distributions for light leptons: 
opposite-sign same-flavor (OSSF) 
(red solid line), opposite-sign different-flavor (OSDF) (green dashed line)
and the OSSF Standard Model background (shaded). 
(Right) Hadronically-decaying ditau mass distributions (including the
$\nu_{\tau}$ four-momenta) with OS (red solid) and SS (green dashed), and the
Standard Model background for OS ditau pairs (shaded). The top
plots are for case A, and the bottom plots are for case B.} 
\label{fig:h3}
\end{figure}

In Fig. \ref{fig:h3} we show (left) dilepton IMDs for light leptons  
and (right) hadronically-decaying $\tau$ leptons after imposing 
the cuts explained in (\ref{eq-cuts}), as fixed in the previous subsection 
\ref{cut-choices}. In the case of point A (top plots), 
both the opposite-sign same-flavour (OSSF) light dileptons (left) and 
$\tau^\pm - \tau^\mp$ distributions (right),
both shown as red solid lines, exhibit breaks at the characteristic 
end-points~\footnote{However, the solid curve in the top right panel of Fig.~\ref{fig:h3} is not observable, 
since the energy carried off by the $\nu_{\tau}$ would escape detection.}, which occur at the invariant mass
\begin{equation}
M_{ll}^{max}=\sqrt{\frac{(M_{\tilde{\chi}_2^0}^2-M_{\tilde{l}}^2)(M_{\tilde{l}}^2-M_{\tilde{\chi}_1^0}^2)}{M_{\tilde{l}}^2}}=98.3 \ \mbox{GeV} .
\end{equation}
On the other hand, the opposite-sign different-flavor (OSDF) (left) and the same-sign
$\tau_{h}^{\pm}\tau_{h}^{\pm}$ distributions (right) do not display such a 
kinematical constraint. 
For point B (lower plots) the OSSF end-point does not exist in the light lepton case, and the peak
at $M_Z$ is completelly obscured by the Standard Model background. In the OS $\tau -\tau$
case, the end-point is at 108.3~GeV.

\subsection{Lepton-Flavour-Violating Events}

\begin{figure}[!hbt]
\begin{center}
\begin{tabular}{cc}
\epsfig{file=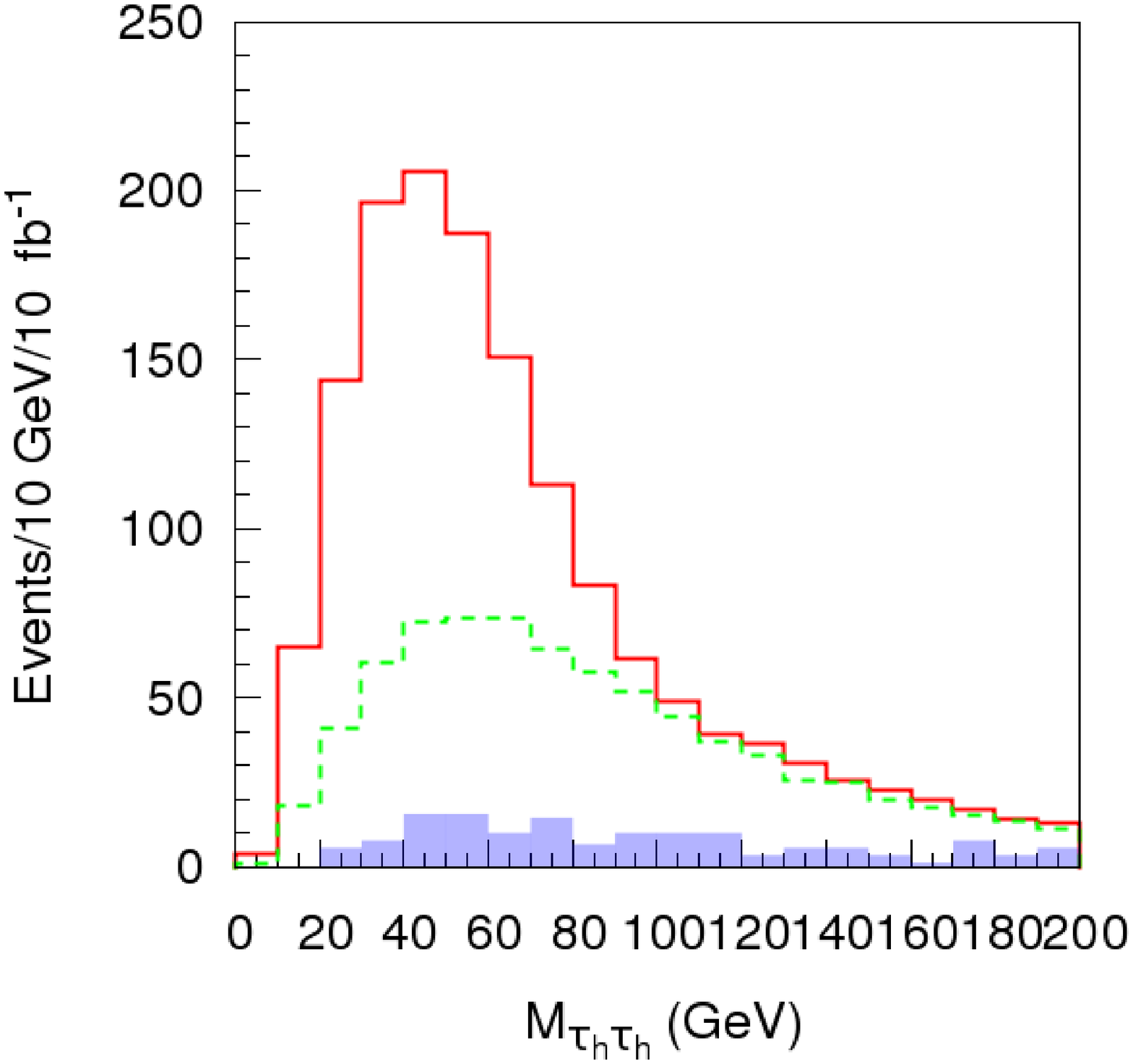,height=2.5in} &
\epsfig{file=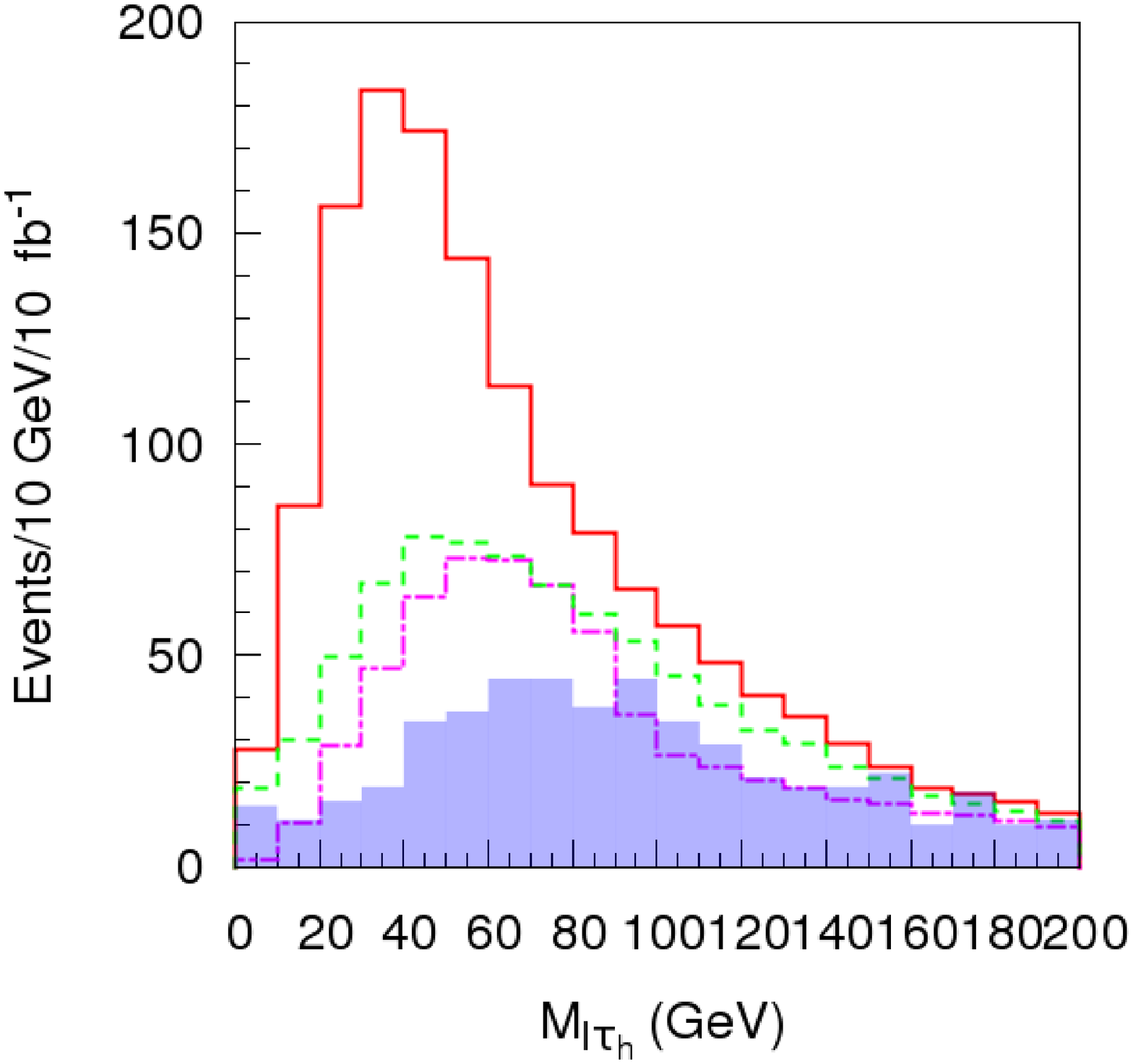,height=2.5in} \\
\epsfig{file=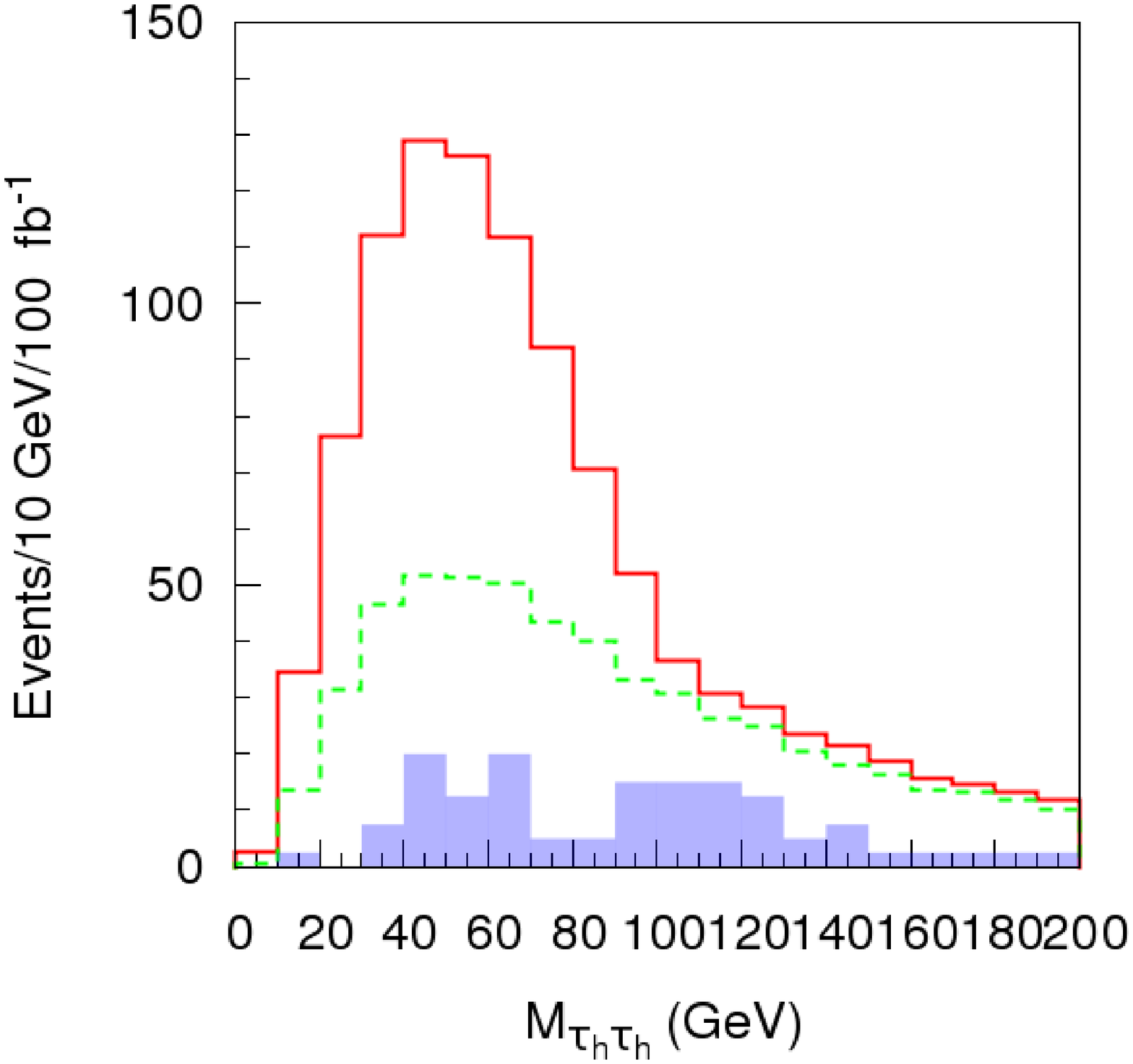,height=2.5in} &
\epsfig{file=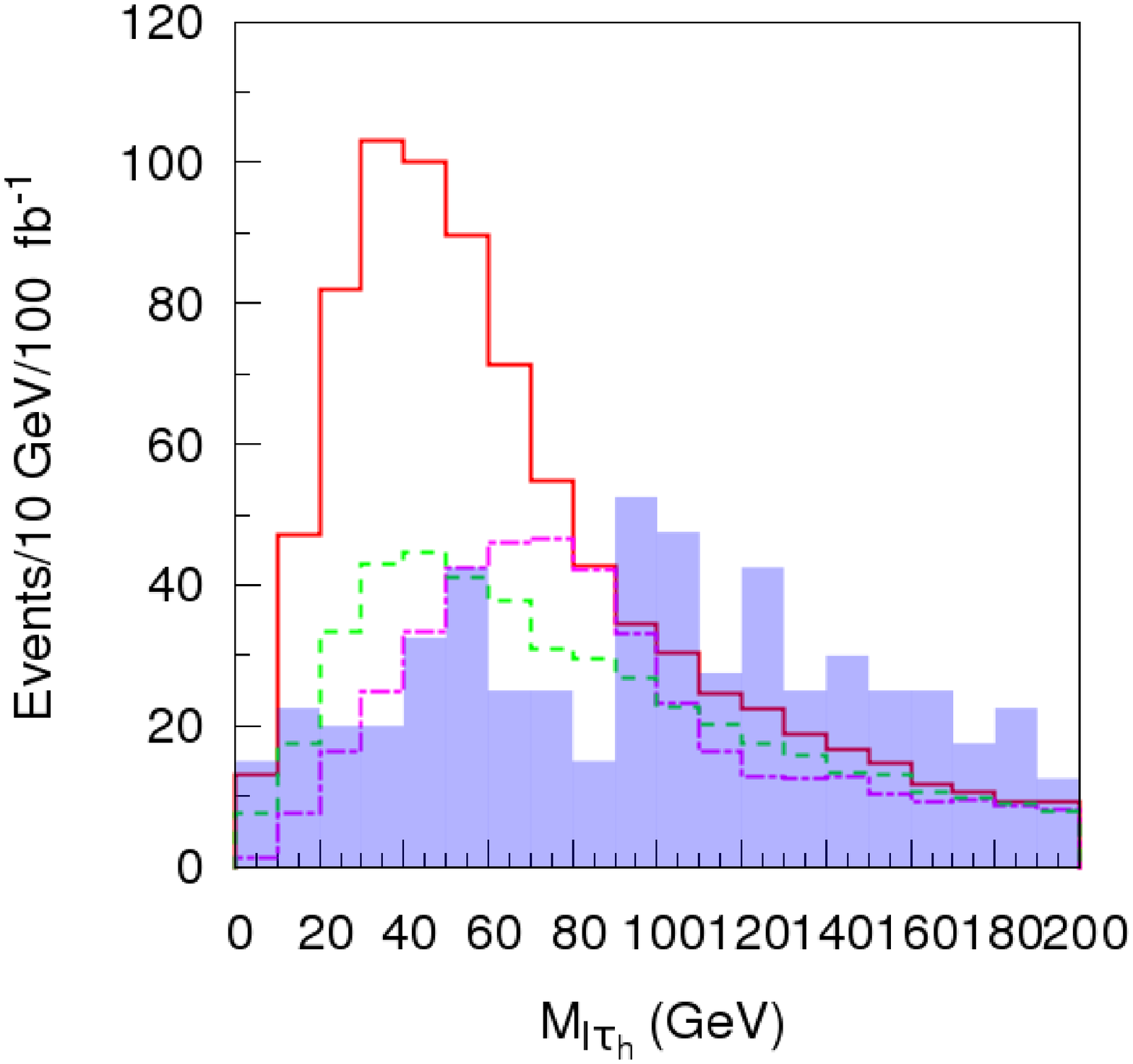,height=2.5in}
\end{tabular}
\end{center}
\caption{\small \it   Left: visible $\tau_h\tau_h$ mass distributions with OS (red solid lines), 
SS (green dashed lines) and the OS Standard Model backgrounds (shaded). 
Right: visible $l\tau_h$ mass distributions with OS (red solid lines), 
SS (green dashed lines), and OS Standard Model backgrounds (shaded).
The LFV OS $\mu\tau_h$ pairs are also shown (pink-dot-dashed). 
As usual, the top plots are for case A and the bottom plots are for case B.} 
\label{fig:h4}
\end{figure}

The left plots of Fig.~\ref{fig:h4} show the visible mass distributions for pairs of hadronically-decaying taus. 
Excesses of OS pairs (red solid lines) over the SS pairs (green dashed lines) ending approximately at the kinematic end-points can be seen.  
The structures of the end-points are not as clean as in Fig.~\ref{fig:h3}, because of the energy 
carried away by neutrinos. 
In the right plots, an excess of OS $l^{\mp}\tau_h^{\pm}$ pairs over the SS pairs can be seen. 
The distributions of OS LFV $\mu\tau$ pairs (that cannot be distinguished 
experimentally from $l^{\mp}\tau_h^{\pm}$)
are also shown for the sake of comparison. 
In view of the difficulty in distinguishing experimentally the LFV $\mu-\tau$ 
signal pairs from the Standard Model background, we
simulate and plot the excess of $\mu-\tau$ pairs over $e-\tau$ pairs. These should
be identical in the Standard Model on average, so any excess of $\mu-\tau$ pairs beyond
statistical fluctuations would be a signal of LFV.

\begin{figure}[!h]
\begin{center}
\begin{tabular}{cc}
\epsfig{file=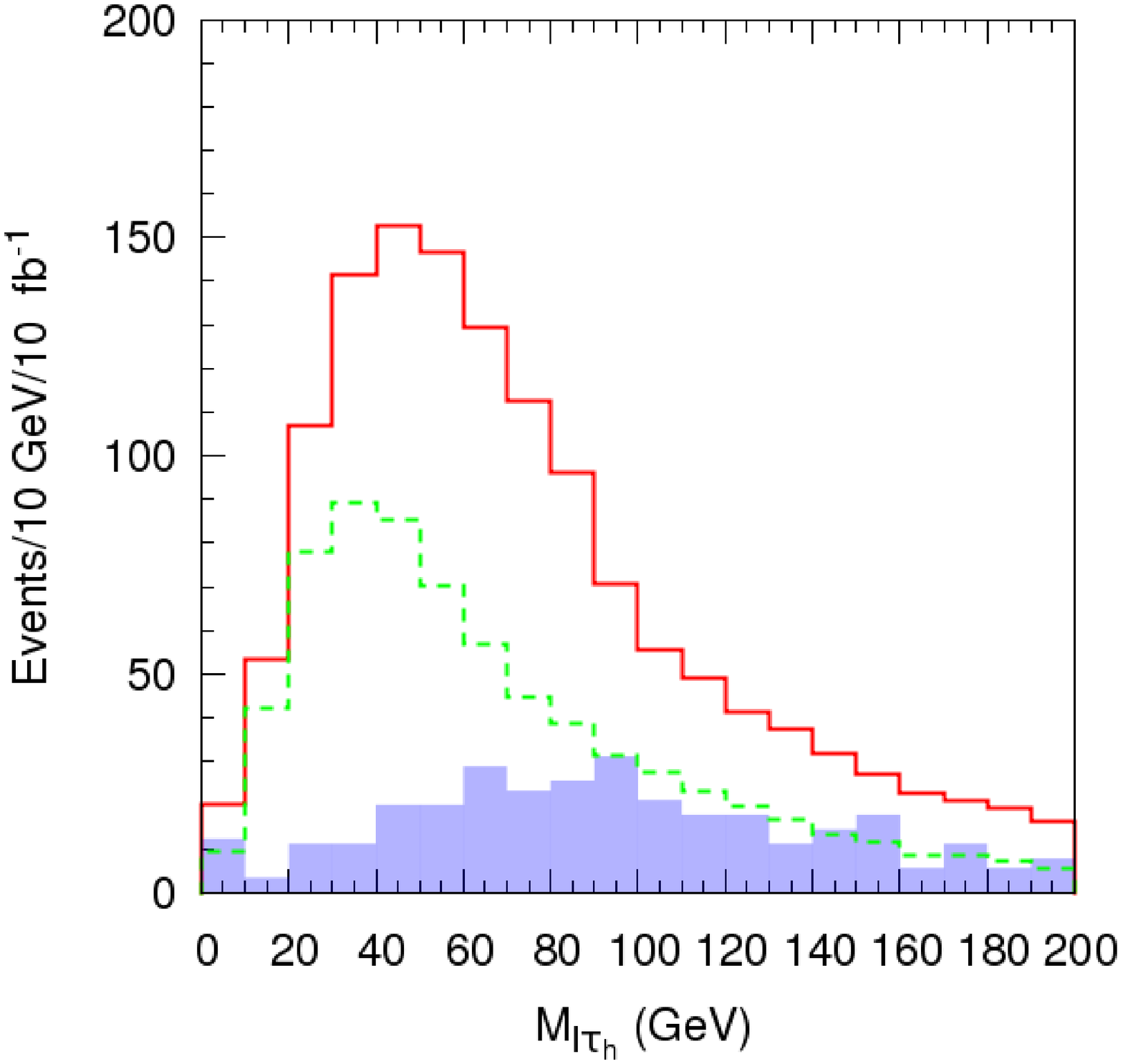,height=2.5 in} &
\epsfig{file=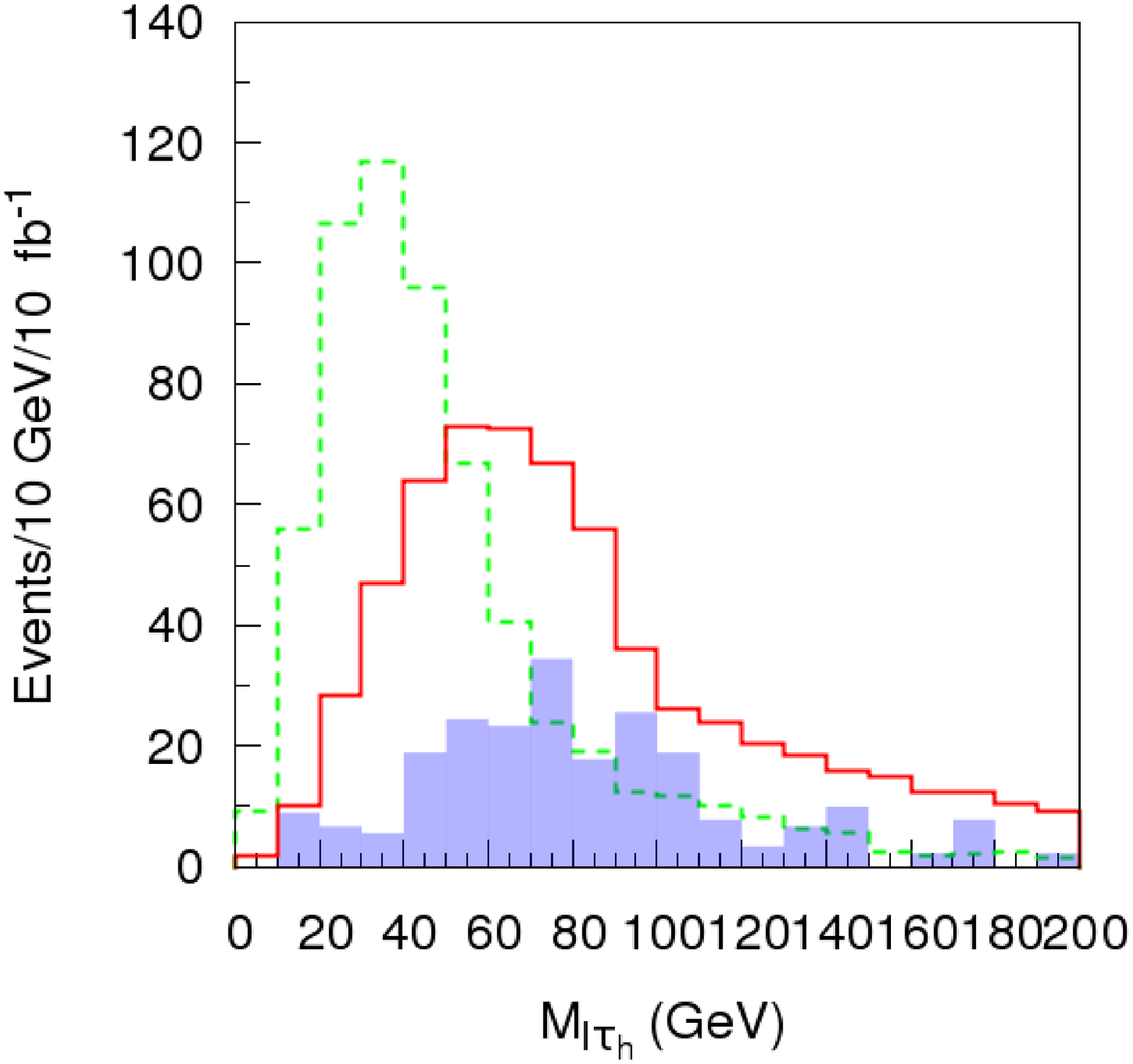,height=2.5 in} \\
\epsfig{file=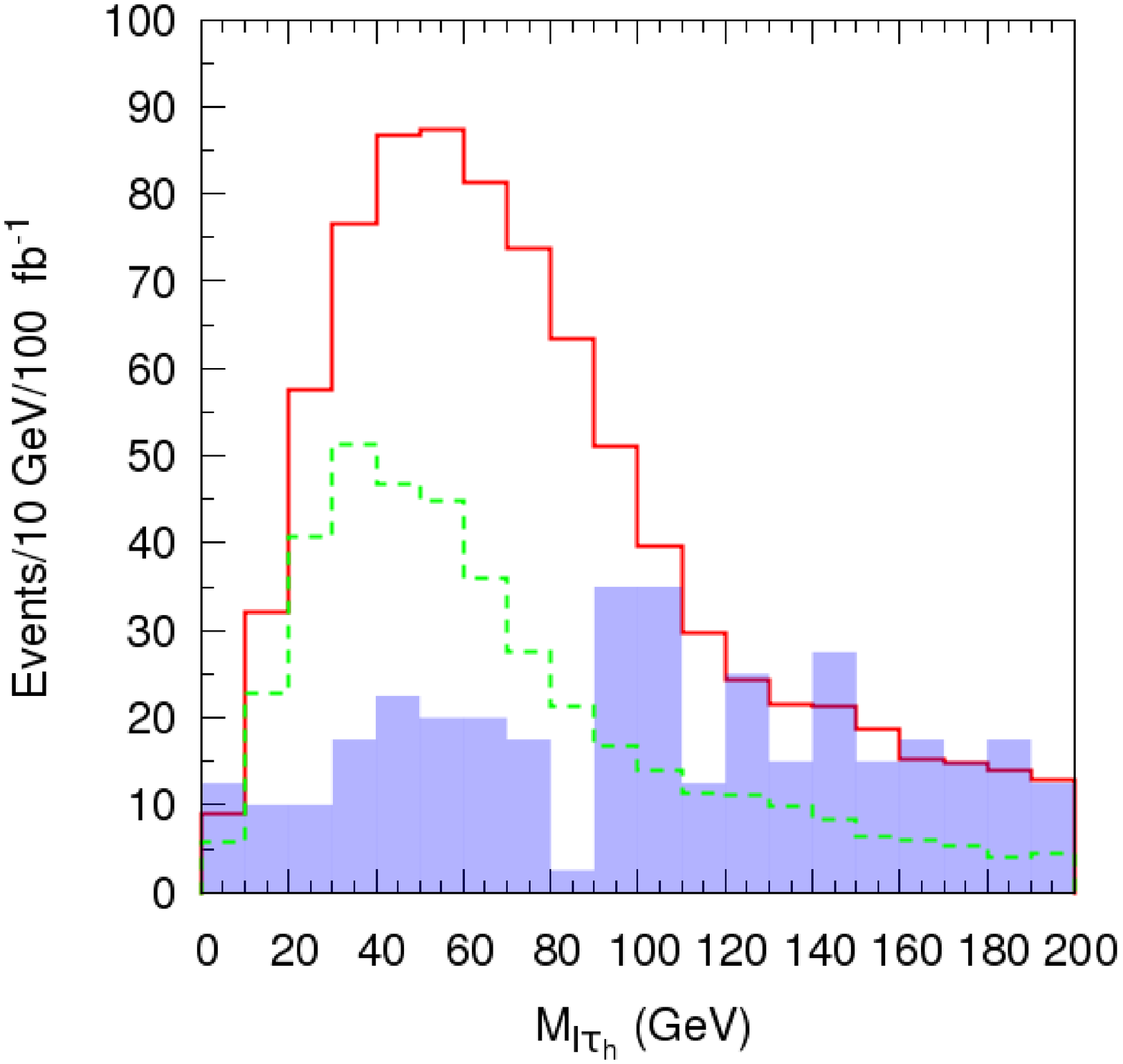,height=2.5 in} &
\epsfig{file=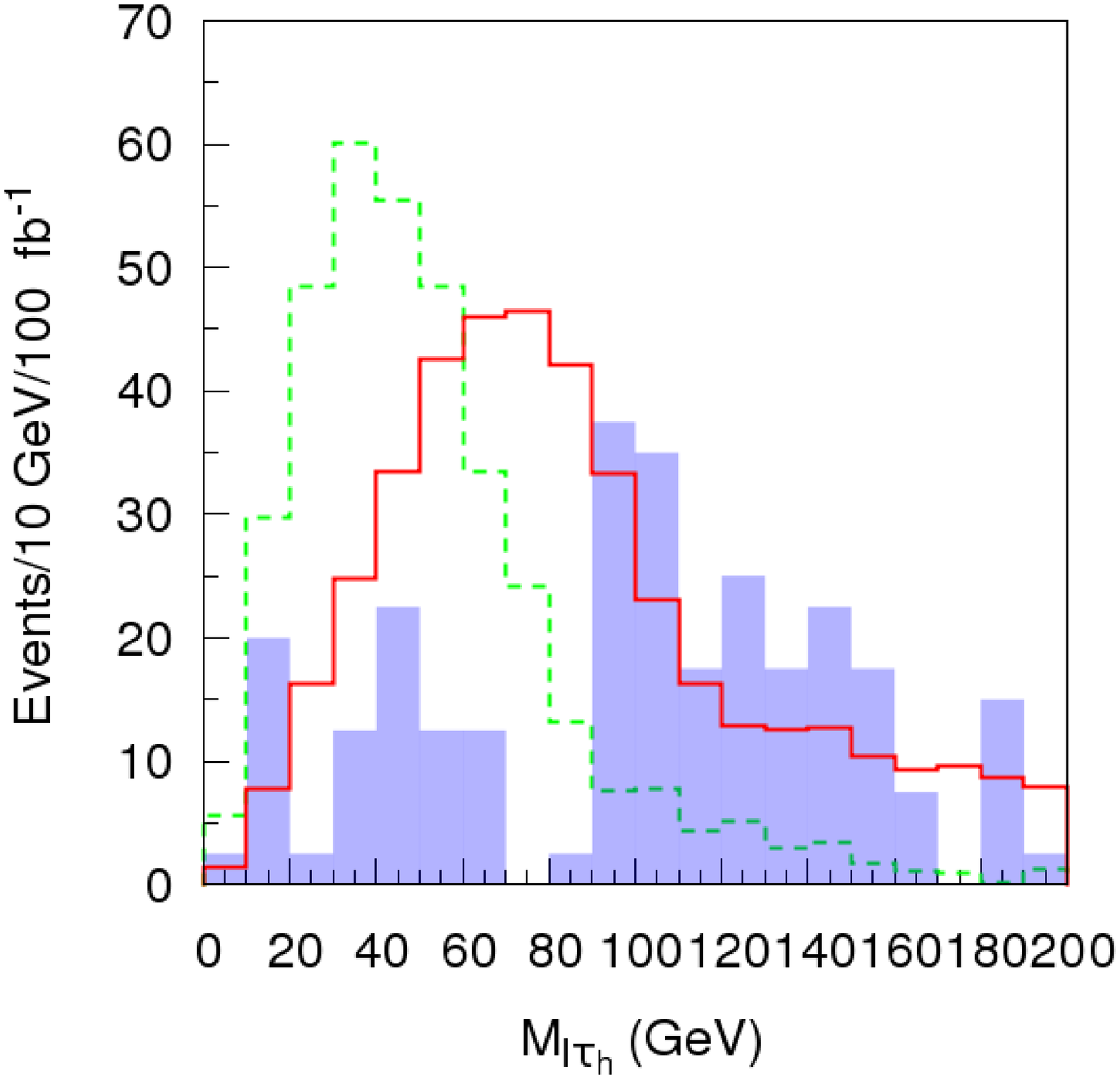,height=2.5 in}\end{tabular}
\end{center}
\caption{\small \it   Left: comparisons between the visible mass distributions for $\tau_h\mu$ (red solid lines), 
$\tau_h e$ (green dashed lines) and Standard Model $\tau_h\mu$ pairs (shaded). 
The LFV $\mu\tau_h$ pairs have been added to the $\mu\tau_h$ distribution.
Right: comparison between the visible mass distributions for LFV $\mu\tau_h$ (red solid lines), 
the sign-subtracted distribution $l^{\pm}\tau_h^{\mp} - l^{\pm}\tau_h^{\pm}$ (green dashed lines), 
and the sign-subtracted Standard Model backgrounds (shaded). } 
\label{fig:h5}
\end{figure}

In fig.~\ref{fig:h5} (left) the excesses of $\mu^{\mp}\tau^{\pm}_h$ over $e^{\mp}\tau^{\pm}_h$ pairs 
for points A and B are shown.  
The observable numbers, $N_{\mu\tau_h}^{lfv}$, of $\mu^{\mp}\tau^{\pm}_h$ LFV pairs are obtained 
by summing the counts in the subtracted
$\mu^{\mp}\tau^{\pm}_h - e^{\mp}\tau^{\pm}_h$ distributions in
the interval of $M_{l\tau}$ masses between $30$ and $110$~GeV. We obtain
\begin{eqnarray}
{\rm Point~A}: \; N_{\mu\tau_h}^{lfv}&=&470 \ \pm 39 \ (12 \ \sigma)
\nonumber \\
 {\rm Point~B}: \; N_{\mu\tau_h}^{lfv}&=&308 \ \pm 30 \ (10 \ \sigma)
\end{eqnarray}
where we quote only the statistical errors for the signal samples. 
If we estimate an efficiency of
70 \% for the jet-tau matching, a lower number of $(\mu\tau_h)_{lfv}$ 
pairs would be obtained:

\begin{eqnarray}
{\rm Point~A}: \; N_{\mu\tau_h}^{lfv} &=& 355 \ \pm 34 \ (10 \ \sigma)
\nonumber  \\
{\rm Point~B}: \; N_{\mu\tau_h}^{lfv} &=& 236 \ \pm 27 \ (9 \ \sigma)
\end{eqnarray}

\begin{figure}[!h]
\begin{center}
\begin{tabular}{cc}
\epsfig{file=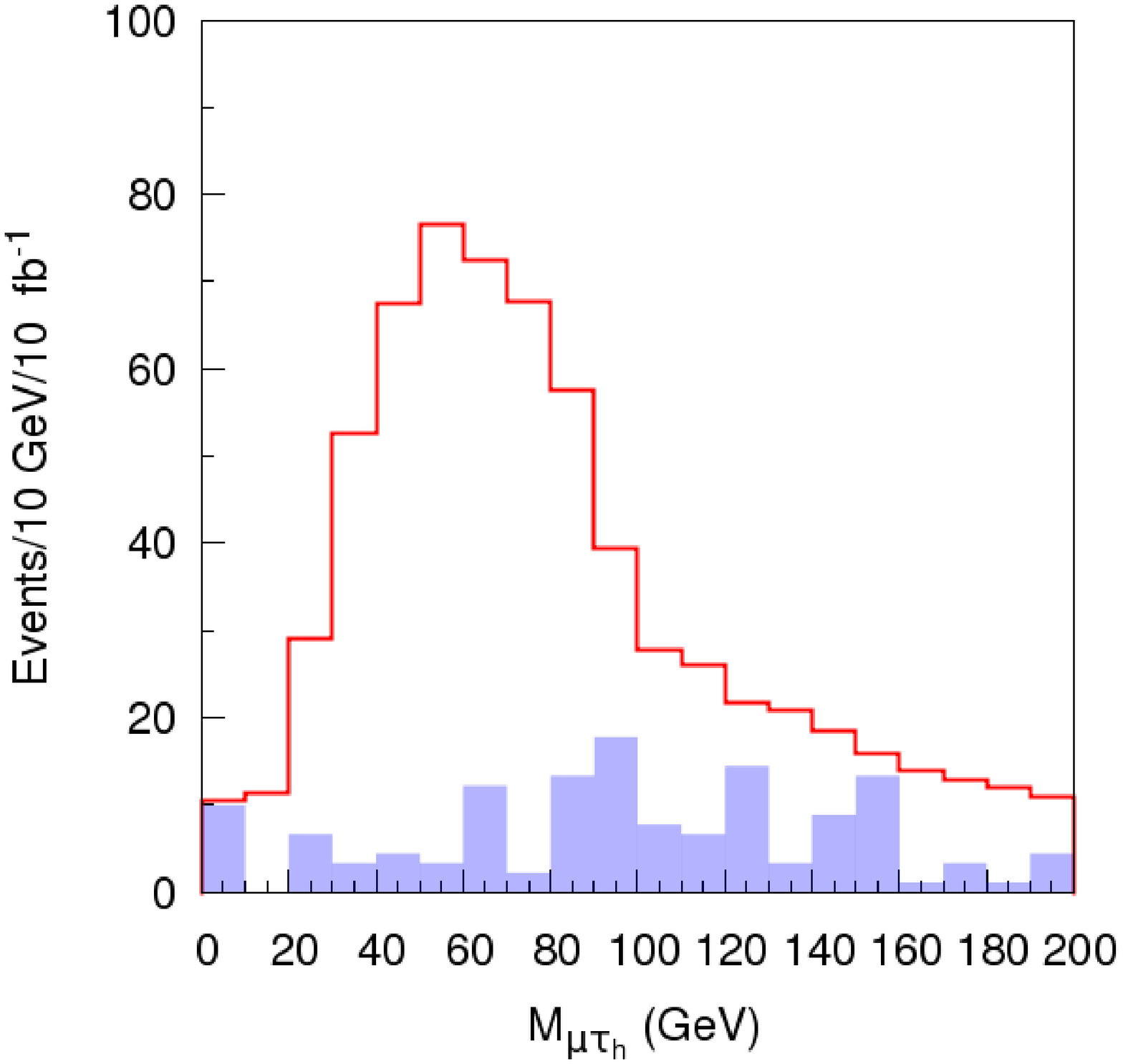,height=2.5 in} &
\epsfig{file=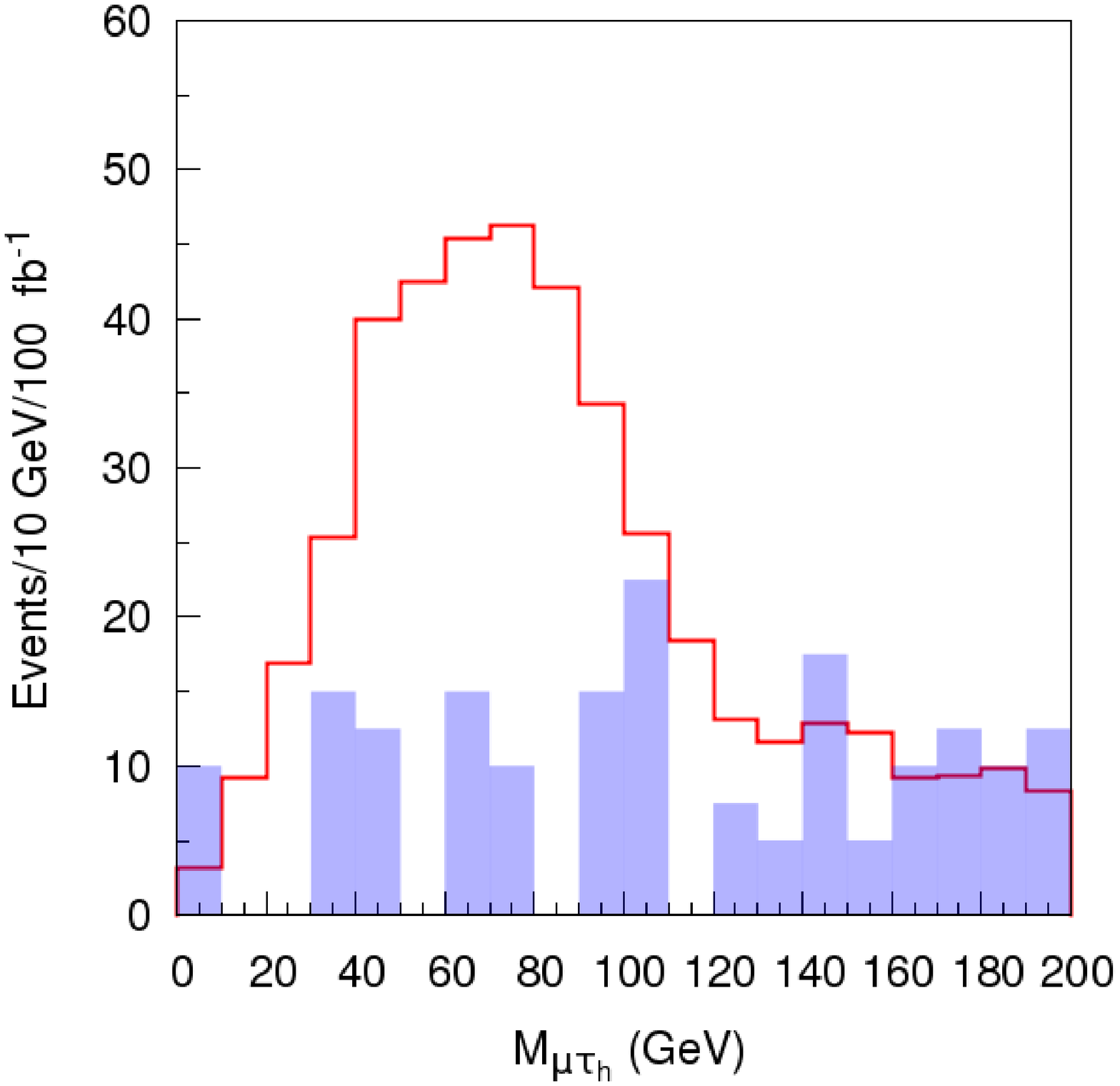,height=2.5 in}
\end{tabular}
\end{center}
\caption{\small \it  The signal for  excess $\mu\tau_h$ LFV pairs (red solid lines) 
and subtracted $\mu\tau_h - e\tau_h$ Standard Model backgrounds (shaded) for points A (left) 
and B (right). As no Standard Model contributions should survive after subtraction, on average,
the shaded signals are due to statistical fluctuations.} 
\label{fig:h6}
\end{figure}

In fig.~\ref{fig:h6}, the excesses of LFV $\mu\tau_h$ pairs over $e\tau_h$ are shown in comparison 
with the subtraction of  $e\tau_h$ pairs from $\mu\tau_h$ pairs  obtained from the simulation of the 
Standard Model background, for points A (left) and B (right). The non-zero subtracted 
signals for the Standard Model result from the statistical fluctuations in our simulation. The
comparison shows that  
the statistical significance of the LFV signal is quite high in both cases, as it can be 
separated well from the Standard Model statistical fluctuation: we obtain $S/B \simeq 2.6$ for 
point B, whilst the ratio is better for point A, namely $S/B \simeq 4.65$.

\subsection{Results for Varying $m_0$ at Fixed $M_{1/2}$}

So far, we have restricted our attention essentially to CMSSM points lying
in the coannihilation strip where the $\tilde \tau_1$ is only slightly heavier than the neutralino
LSP, which restricts the kinematics of the $\tau$ hadronic jet. Points with the same
value of $M_{1/2}$ but larger values of $m_0$ may also be allowed under certain
circumstances, e.g., in direct-channel resonance regions that may appear at large $\tan \beta$ in the 
CMSSM or at lower $\tan \beta$ in models with non-universal Higgs masses, or if the gravitino is the 
LSP. Points with larger $m_0$ have different kinematics, so here we study the consequences for
the observability of LFV in such models.
We fix $M_{1/2} = 500$~GeV and $\tan \beta = 35$, and vary $m_0$ and (slightly) $A_0$.
We choose four reference points, whose parameters
are listed in Table~5, and the resulting sparticle mass spectra in Table~6.

\begin{table}[!h]
\begin{center}
\begin{tabular}{| c | c | c | c | c | c | c | c |}
\hline
Point & $m_0$ & $M_{1/2}$ & $\tan\beta$ & $A_0$ & N$_{events}$ & $\sigma_{int}$ & $L_{int}$ \\
\hline
\hline
1 & 250 & 500 & 35 & 250 & 517K & 1.72~pb & 300~fb$^{-1}$ \\
\hline
2 & 300 & 500 & 35 & 300 & 494K & 1.65~pb & 300~fb$^{-1}$ \\
\hline
3 & 350 & 500 & 35 & 350 & 470K & 1.57~pb & 300~fb$^{-1}$ \\
\hline
4 & 400 & 500 & 35 & 400 & 442K & 1.48~pb & 300~fb$^{-1}$\\ 
\hline 
\end{tabular}

\caption{\small \it   Parameters of the four CMSSM reference points 1, 2, 3, 4 with increasing
values of $m_0$ and $A_0$ (all mass parameters are given in GeV units). 
We also quote the numbers of events, the LHC 
cross sections and the assumed sample luminosities.}

\end{center}
\label{table2}
\end{table} 

\begin{table}[!h]
\begin{center}
\begin{tabular}{| c | c | c | c | c | c | c | c | c | c |}
\hline
Point & $M_{\tilde{g}}$ & $M_{\tilde{u}_L}$ & $M_{\tilde{d}_L}$ & $M_{\tilde{\chi}_2^0}$ & $M_{\tilde{\tau}_1}$ & $M_{\tilde{\chi}_1^0}$ & $M_{\tilde{l}_R}$ & $M_{\tilde{l}_L}$ & $M_{h}$ \\
\hline
\hline
1 & $1155$ & $1080$ & $1084$ & $388$ & $242$ & $206$ & $314$ & $422$ & $116$ \\
\hline
2 & $1157$ & $1092$ & $1095$ & $388$ & $282$ & $206$ & $355$ & $453$ & $115$ \\
\hline
3 & $1159$ & $1106$ & $1109$ & $388$ & $323$ & $206$ & $398$ & $487$ & $115$\\ 
\hline
4 & $1162$ & $1123$ & $1126$ & $388$ & $363$ & $206$ & $442$ & $524$ & $115$\\
\hline
\end{tabular}
\caption{\small \it   Relevant sparticle masses (in GeV) for the reference points defined in 
Table~5.}
\label{table3}
\end{center}
\end{table}

\begin{figure}[!h]
\begin{center}
\hspace*{-.5in}
\epsfig{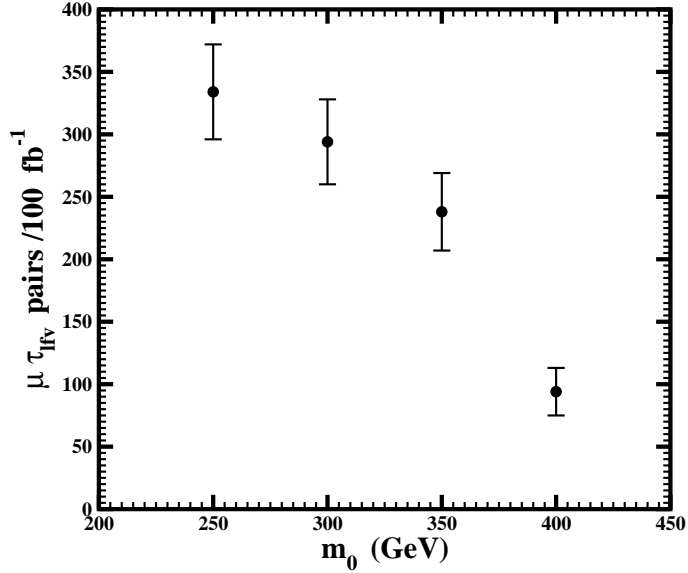} 
\caption{\small \it    Observable numbers of LFV $\mu^{\mp}\tau^{\pm}_h$ pairs for the different points
described in Tables 5 and 6, with increasing values of $m_0$ and $A_0$, for fixed $M_{1/2}
= 500$~GeV and $\tan \beta = 35$.}
\label{fig:pts} 
\end{center}
\end{figure}

We display in Fig.~\ref{fig:pts} the numbers of LFV $\mu^{\mp}\tau^{\pm}_h$ pairs
for each of these points, calculated in the same way as in previous subsection, for the points 
described in Tables 5 and 6. We also display the statistical error bars. We see that the
LFV signal continues to be observable up to the largest value of $m_0$ studied, namely
400~GeV for point~4 above. We conclude that the analysis described previously in this 
paper is quite robust, and the LFV signal has a good likelihood of being observable, as
long as its branching ratio exceeds about 10\%.

\section{Conclusions}

Motivated by the neutrino oscillation data, and the ensuing likelihood that low-energy
violations of charged lepton numbers might be detectable in supersymmetric seesaw
models, we have explored the observability at the LHC of sparticle decays that violate $\tau$ 
lepton number. We have worked in the context of the Constrained Minimal Supersymmetric 
extension of the Standard Model (CMSSM) and in a non-minimal SU(5) GUT extension of the 
theory. Focusing mainly on regions of the CMSSM parameters with values of the relic neutralino 
LSP density that fall within the range acceptable to cosmology, we investigate have the SUSY 
parameter space requirements for tau flavour violation to be observable in 
$\chi_2 \to \chi + \tau^\pm \mu^\mp$ decays. We have studied the possible
signals from hadronic $\tau$ decays, which we have analyzed at the event
generator level with the use of {\tt PYTHIA}.

Within this framework, we have found the following:
\begin{itemize}
\item The observation of LFV in neutralino decays at the LHC can 
be possible if ${\Gamma(\chi_{2}\rightarrow\chi_{1}\tau^{\pm}\mu^{\mp})}$
$/{\Gamma(\chi_{2}\rightarrow\chi_{1}\tau^{\pm}\tau^{\mp})}\sim 0.1$.
\item The strong bounds on radiative $\tau$ decays, as well as the other cosmological and
phenomenological requirements, constrain significantly the allowed parameter space,
to the extent that the CMSSM and the minimal SU(5) GUT are
not promising frameworks for observing LFV sparticle decays.
\item
Larger ratios can be found in a non-minimal SU(5), where $RR$ slepton mixing may be
substantial, enabling the LFV signal to be distinguished 
clearly from the background.
\item
The LFV signal remains observable also at larger values of $m_0$ than are favoured
in the usual CMSSM framework.
\end{itemize}
We conclude that the search for $\chi_{2}\rightarrow\chi_{1}\tau^{\pm}\mu^{\mp}$
decays at the LHC is interesting and complementary to the parallel searches for
$\tau \to \mu \gamma$ decays, and could be a useful `canary in the mine' for
non-minimal GUTs.


\vskip 1. cm
~\\
{\bf Acknowledgements} 

E. Carquin and S. Lola thank the CERN Theory Division and the A.P.
Department of the University of Huelva for kind hospitality during several
phases of this work.  The work of E. Carquin has been partly supported  by the
MECESUP Chile Grant
and the HELEN Program. The work of M.E.G and J.R.Q is supported by the
Spanish MEC project FPA2006-13825 and the project P07FQM02962 funded
by ``Junta de Andalucia''.
The research of  S. Lola is funded by the FP6
Marie Curie Excellence Grant MEXT-CT-2004-014297.

\vskip 1. cm


\end{document}